\documentclass[runningheads,citeauthoryear]{apinv}
\usepackage{epsfig,cite,graphics}
\usepackage[T2A]{fontenc}
\usepackage[cp1251]{inputenc}
\usepackage{lscape}
\usepackage{longtable}

\begin{document}

\title{Photometric variability of 14 PMS stars in the NGC 7000/IC 5070 complex}
\titlerunning{Photometric variability of 14 PMS stars in the NGC 7000/IC 5070 complex}
\author{Sunay I. Ibryamov\inst{1,2} and Evgeni H. Semkov\inst{1}}
\authorrunning{S. Ibryamov and E. Semkov}
\tocauthor{S. Ibryamov and E. Semkov} 
\institute{Institute of Astronomy and NAO, Bulgarian Academy of Sciences, BG-1784 Sofia
	\and Department of Theoretical and Applied Physics, University of Shumen, BG-9712 Shumen
	\newline
	\email{sibryamov@astro.bas.bg}}
\papertype{Submitted on; Accepted on}	
\maketitle

\begin{abstract}
New photometric data from CCD multicolour $BVRI$ observations of 14 pre-main sequence stars during the period from 2013 April to 2015 September are presented. The studied objects are located in the field of 'Gulf of Mexico' in the NGC 7000/IC 5070 star-forming complex. The stars from our study exhibit different types of photometric variability in all optical passbands. Using our long-term observations and data published by other authors, we tried to define the reasons for the observed brightness variations. On the basis of our new data previously unknown periodicity in the light curve of the star LkH$\alpha$ 189 (2.45 days) was registered. 
\end{abstract}


\section*{1. Introduction}

Studies of pre-main sequence (PMS) stars give an opportunity to understand the early stages of stellar evolution and to test stellar evolution scenarios. Depending on their initial mass, PMS stars pass through different periods of stellar activity. The most prominent manifestations of this activity are changes in the star's brightness with various periods and amplitudes.

PMS stars are rare among field stars, because during the PMS evolutionary phase the stars spend less then 1\% of its life. PMS stars are separated into two main types, low mass (M $\leq$ 2M$_{\sun}$) T Tauri stars (TTS) and the more massive (2M$_{\sun}$ $\leq$ M $\leq$ 8M$_{\sun}$) Herbig Ae/Be stars (HAEBES). Both TTS and HAEBES still contracting towards the main sequence.

TTS are associated with dark nebulae, molecular clouds, obscured regions, and are grouped in T associations. The main characteristics of TTS are their strong irregular photometric variability and emission line spectra (Joy 1945). TTS are divided into two sub-classes: classical T Tauri stars (CTTS) and weak-line T Tauri stars (WTTS). Most of the features that characterize each sub-class suggest that accretion disks surround CTTS whereas they must have almost disappeared (at least their inner parts) in WTTS (see Bertout 1989; M\'{e}nard \& Bertout 1999).

According to Herbst et al. (2007), variability of CTTS can be due to rotating hot spots on the stellar surface. CTTS also often shows irregular variations with large photometric amplitudes up to 2-3 mag, associated with high variable accretion from the circumstellar disk onto the stellar surface. The variability of WTTS is due to cool spots or groups of spots on the stellar surface. The photometric amplitudes of this variability are about 0.03-0.3 mag, but in extreme cases reach to 0.8 mag in the $V$-band. The variability of WTTS can be due to flare-like variations in $B$- and $U$-band also. Flares are random with different sized amplitudes, as there is no periodicity. Usually, the rise of brightness is quick and occurs over short time scales.

The large amplitude drops in the brightness in the light curves are commonly observed in the early types (G, F) of TTS and HAEBES. These stars show non-periodic deep Algol-type minima and photometric amplitudes up to 2.8 mag in $V$-band. The observed drops in the brightness last from days to some weeks and are caused by variable extinction from circumstellar dust or clouds (Voshchinnikov 1989; Grinin et al. 1991; Herbst et al. 2007). This group of PMS stars with intermediate mass is known as UXors and their prototype is the star UX Orionis. In very deep minima, the colour indices of UXors often becomes bluer (so called 'blueing effect' or 'colour reverse') (see Bibo \& Th\'{e} 1990). It is generally accepted that the origin of observed drops in the brightness and the blueing effect are due to variations of the colomn density of dust in the line of sight to the star (Dullemond et al. 2003).

The stars from our study are located in the dense molecular cloud L935, known as 'Gulf of Mexico' in the NGC 7000/IC 5070 complex. The NGC 7000/IC 5070 complex are thought to be parts of a single large HII region W80. The distance to this region, determined by Laugalys \& Strai\v{z}ys (2002) is 600 pc. 'Gulf of Mexico' is a region with active star formation and contains many PMS objects (see Armond et al. 2011; Findeisen et al. 2013; Bally et al. 2014; Poljancic Beljan et al. 2014). The PMS stars included in the present study are revealed in previous studies of the 'Gulf of Mexico' as H$\alpha$ emission stars, flare stars from UV Cet type, TTS or HAEBES.

The present paper is a part of our long-term multicolour photometric study of the PMS stars in the region of 'Gulf of Mexico'. The results from our previous studies of PMS stars in this field have been published in Semkov et al. (2010, 2012, 2014), Poljan\v{c}i\'{c} Beljan et al. (2014), Ibryamov et al. (2015a), and Ibryamov et al. (2015b).
Section 2 gives information about telescopes and cameras used and data reduction. Section 3 describes the obtained results and their interpretation.

\section*{2. Observations and Data reduction}

The presented photometric data were carried out in the period from 2013 April to 2015 September. The observations were performed with four telescopes: 2- Ritchey-Chr\'{e}tien-Coud\'{e} (RCC), the 50/70-cm Schmidt and the 60-cm Cassegrain telescopes of the Rozhen National Astronomical Observatory in Bulgaria and the 1.3-m Ritchey-Chr\'{e}tien (RC) telescope of the Skinakas Observatory\footnote{Skinakas Observatory is a collaborative project of the University of Crete, the Foundation for Research and Technology, Greece, and the Max-Planck-Institut f{\"u}r Extraterrestrische Physik, Germany.} of the University of Crete in Greece. The total number of the nights used for observations is 84. A log of the observing periods and telescopes used is given in Table 1.

\begin{table}[htb!]
  \caption{Photometric CCD observations of the field of 'Gulf of Mexico' with the different telescopes}
  \begin{center}
  \begin{tabular}{lcc}
	  \hline\hline
	  \noalign{\smallskip}
Telescope  & Nr of observations & Time span \\
   \noalign{\smallskip}
   \hline
   \noalign{\smallskip}
2-m RCC           & 23 & 04.07.2013 $-$ 06.09.2015 \\
1.3-m RC          &  4 & 17.09.2013 $-$ 12.08.2015 \\
50/70-cm Schmidt  & 41 & 10.04.2013 $-$ 03.09.2015 \\
60-cm Cassegrain  & 16 & 15.05.2013 $-$ 21.07.2014 \\
\hline \hline
  \end{tabular}
  \end{center}
  \end{table}

The observations were performed with four different types of CCD cameras: VersArray 1300B at the 2-m RCC telescope, ANDOR DZ436-BV at the 1.3-m RC telescope, FLI PL16803 at the 50/70-cm Schmidt telescope, and FLI PL09000 at the 60-cm Cassegrain telescope. The technical parameters and specifications for the cameras used, observational procedure, and data reduction process are given in Ibryamov et al. (2015b).
The photometric limit of our data obtained with the 2-m RCC and the 1.3-m RC telescopes in $V$- and $B$-band is about 20.5 mag and for the data collected with the 50/70-cm Schmidt and the 60-cm Cassegrain telescopes is about 19.5 mag.

All frames were taken through a standard Johnson-Cousins ($BVR_{c}I_{c}$) set of filters. All data were analyzed using the same aperture, which was chosen to have a 4$\arcsec$ radius, while the background annulus was taken from 9$\arcsec$ to 14$\arcsec$. As a reference, the $BVRI$ comparison sequence reported in Semkov et al. (2010) was used.
The mean value of the errors in the reported magnitudes are 0.01-0.02 mag for $I$- and $R$-band data, 0.02-0.06 mag for $V$-band data, and 0.02-0.09 mag for $B$-band data.

\section*{3. Results and Discussion}

The photometric results for 14 stars obtained in our study and their interpretation are presented in the section. All stars exhibit photometric variability in $BVRI$ optical passbands. Our data are a continuation of the long-term photometric investigations of these stars published in Poljan\v{c}i\'{c} Beljan et al. (2014) and Ibryamov et al. (2015a).

The data for the studied objects are presented in Table 2. Star identifiers used in this paper are marked in boldface. Figure 1 shows a three-colour images of the field of 'Gulf of Mexico', where the positions of the stars from our study and the recently erupting FUor star V2493 Cyg (see Semkov et al. 2010, 2012, 2014; Miller et al. 2011) are marked. The images were obtained with the 50/70-cm Schmidt telescope of Rozhen NAO on 2012 September 04 (the left image) and on 2014 August 19 (the right image).

The stars from our study shows bursts and/or fades events in their light curves. These events are defined as follows. In the case that the star most of the time spends at high light during our study, the declines in the brightness are interpreted as fades. Respectively, if the star most of the time spends at a low light, the observed increases in the brightness are interpreted as bursts.

	\begin{landscape}
\begin{figure}[]
  \begin{center}
    \centering{\epsfig{file=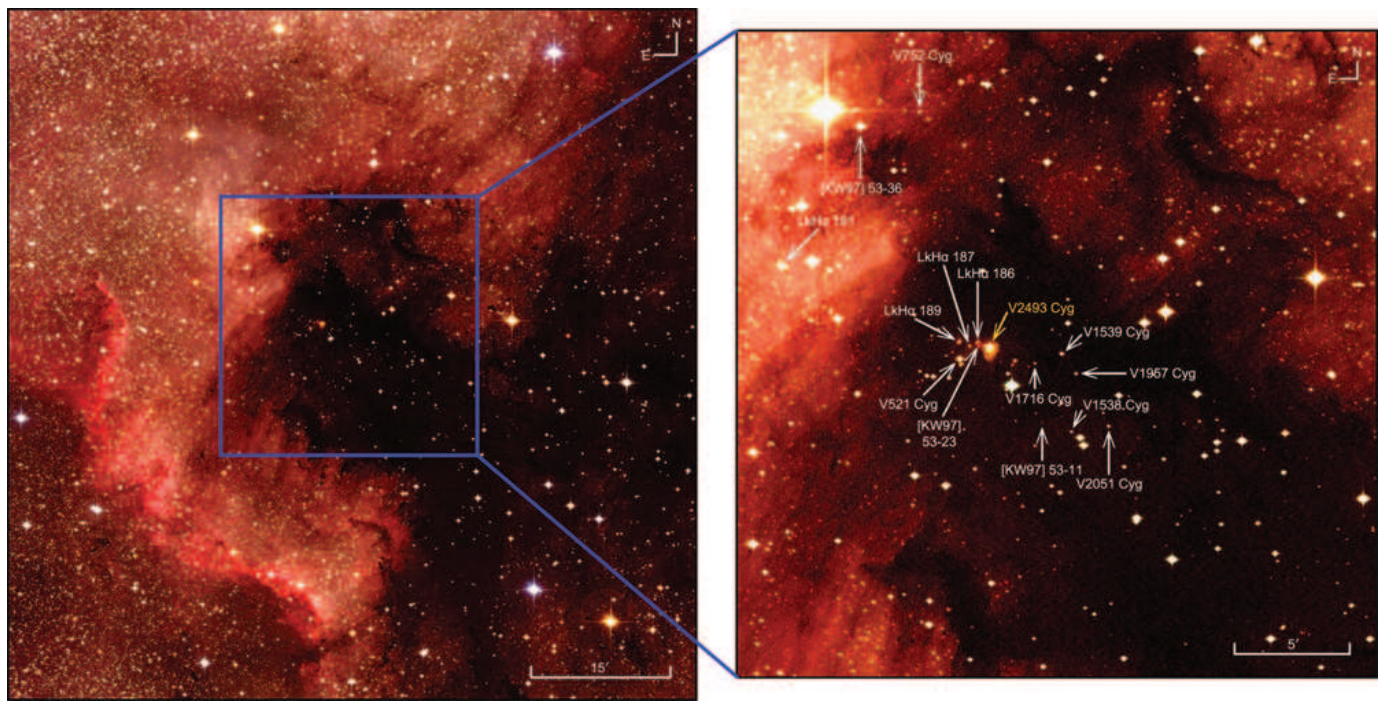, width=18.0cm}}
    \caption[]{Three-colour images of the field of 'Gulf of Mexico'. The positions of the stars from our study and the recently erupting FUor star V2493 Cyg are marked. The images were obtained with the 50/70-cm Schmidt telescope of Rozhen NAO on 2012 September 04 (the left image) and on 2014 August 19 (the right image).}
    \label{fig1}
  \end{center}
\end{figure}
\end{landscape}

The $JHK_{s}$ 2MASS magnitudes of the studied stars were used to construct the two-colour diagram and to identify stars with infrared excess, indicating the presence of a disk. Fig. 2 shows the location of main sequence (brown belt) and giant stars (orange belt) from Bessell \& Brett (1988), and the classical T Tauri stars location (green belt) from Meyer et al. (1997). A correction to the 2MASS photometric system was performed following the prescription of Carpenter (2001). The three parallel dotted lines show the direction of the interstellar reddening vectors determined for the 'Gulf of Mexico' region by Strai\v{z}ys et al. (2008). In Fig. 2 the stars are designated using the numbers from Table 2. The stars lying to the right of the middle vector indicate the presence of a infrared excess. Due to the photometric variability, the positions of stars in Fig. 2 may vary around the displayed values.

Table 3 contain the mean photometrical magnitudes and colour indices for the stars from our study.

{\footnotesize
\begin{table}[]
  \caption{Designations and coordinates of the stars from our study}
  \begin{center}
  \begin{tabular}{llcrcccc}
	  \hline\hline
	  \noalign{\smallskip}
Nr  & GCVS$^1$ & HBC$^2$ & [KW97]$^3$ & LkH$\alpha$ $^4$ & 2MASS ID$^5$ & $RA_{J2000.0}$ & $Dec_{J2000.0}$\\
   \noalign{\smallskip}
   \hline
   \noalign{\smallskip}
1  & \textbf{V521 Cyg}  & 299 & 53-26 & 188  & J20582381+4353114 & 20 58 23.81 & +43 53 11.45\\  
2  & \textbf{V752 Cyg}  &     & 53-30 &      & J20583359+4403354 & 20 58 33.60 & +44 03 35.50\\   
3  & \textbf{V1538 Cyg} &     &       &      & J20575750+4350089 & 20 57 57.51 & +43 50 08.95\\
4  & \textbf{V1539 Cyg} & 720 & 53-9  & 185  & J20575986+4353260 & 20 57 59.86 & +43 53 26.00\\
5  & \textbf{V1716 Cyg} &     &       &      & J20580611+4353011 & 20 58 06.12 & +43 53 01.17\\
6  & \textbf{V1957 Cyg} &     &       &      & J20575651+4352362 & 20 57 56.52 & +43 52 36.29\\
7  & \textbf{V2051 Cyg} &     &       &      & J20574880+4350236 & 20 57 48.80 & +43 50 23.60\\      
8  &      & 723 & 53-24 & \textbf{186}       & J20581961+4353545 & 20 58 19.62 & +43 53 54.52\\
9  &      & 724 & 53-25 & \textbf{187}       & J20582154+4353449 & 20 58 21.54 & +43 53 44.96\\
10 &      & 725 & 53-27 & \textbf{189}       & J20582400+4353546 & 20 58 24.01 & +43 53 54.61\\          
11 &      & 301 & 53-42 & \textbf{191}       & J20590581+4357030 & 20 59 05.82 & +43 57 03.08\\    
12 &  		&     & \textbf{53-11}      &      &                   & 20 58 03.00 & +43 50 42.00\\
13 &      &     & \textbf{53-23}      &      & J20581951+4353449 & 20 58 19.52 & +43 53 45.00\\
14 &      &     & \textbf{53-36}      &      & J20584756+4402574 & 20 58 47.52 & +44 02 59.29\\
   	  \hline \hline
  \end{tabular}
  \end{center}
  {$^1$General Catalogue of Variable Stars (Samus et al. 2009); $^2$Herbig \& Bell (1988); $^3$Kohoutek \& Wehmeyer (1997); $^4$Herbig (1958); $^5$2Micron All-Sky Survey (Skrutskie et al. 2006)}.
  \end{table}} 
	
\begin{figure}[]
  \begin{center}
    \centering{\epsfig{file=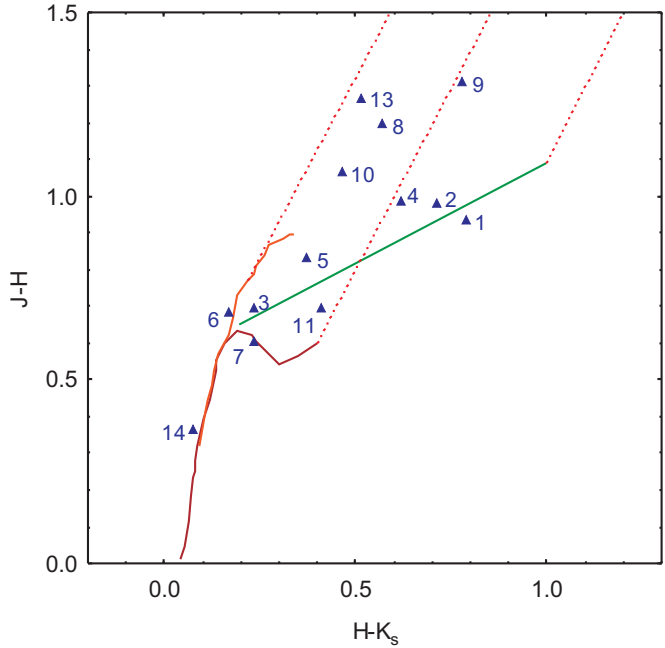, width=9.0cm}}
    \caption[]{The $J-H$ versus $H-K_{s}$ diagram for the stars detected in all 3 bands in 2MASS catalogue.}
    \label{fig2}
  \end{center}
\end{figure}

\begin{table}[]
  \caption{Mean photometrical magnitudes and colour indices for the stars from our study}
  \begin{center}
  \begin{tabular}{lccccccc}
	  \hline\hline
	  \noalign{\smallskip}
Star & $\overline{I}$ & $\overline{R}$ & $\overline{V}$ & $\overline{B}$ & $\overline{V-I}$ & $\overline{V-R}$ & $\overline{B-V}$ \\
   \noalign{\smallskip}
   \hline
   \noalign{\smallskip}
V521 Cyg        & 12.11 & 12.93 & 13.76 & 15.03 & 1.66 & 0.84 & 1.26 \\
V752 Cyg        & 14.98 & 15.89 & 16.61 & 17.48 & 1.63 & 0.72 & 0.87 \\
V1538 Cyg       & 15.00 & 16.04 & 17.00 & 18.38 & 2.00 & 0.95 & 1.38 \\
V1539 Cyg       & 13.51 & 14.51 & 15.55 & 16.90 & 2.04 & 1.04 & 1.35 \\
V1716 Cyg       & 15.01 & 16.36 & 17.51 & 19.07 & 2.51 & 1.16 & 1.56 \\
V1957 Cyg       & 14.57 & 15.43 & 16.26 & 17.50 & 1.70 & 0.83 & 1.24 \\
V2051 Cyg       & 13.93 & 15.56 & 16.61 & 18.12 & 2.68 & 1.05 & 1.52 \\
LkH$\alpha$ 186 & 14.60 & 16.07 & 17.38 & 19.00 & 2.77 & 1.31 & 1.63 \\
LkH$\alpha$ 187 & 14.72 & 16.23 & 17.61 & 19.26 & 2.90 & 1.38 & 1.67 \\
LkH$\alpha$ 189 & 13.97 & 15.29 & 16.54 & 18.21 & 2.57 & 1.25 & 1.67 \\
LkH$\alpha$ 191 & 11.58 & 12.19 & 12.87 & 13.98 & 1.29 & 0.68 & 1.11 \\
$[$KW97$]$ 53-11$^*$& 16.18 & 18.65 & -     & -     & -    & -    & -    \\
$[$KW97$]$ 53-23    & 14.99 & 16.92 & 18.55 & 20.33 & 3.57 & 1.66 & $~$2.01\\
$[$KW97$]$ 53-36    & 11.63 & 11.88 & 12.28 & 13.05 & 0.65 & 0.40 & 0.78 \\
   	  \hline \hline
  \end{tabular}
  \end{center}
  {$^*$ $\overline{R-I}$=2.54}.
  \end{table}

\subsection*{3.1. V521 Cyg} 

Variability of V521 Cyg was discovered and the star was classified as a RW Aurigae type by Hoffmeister (1949). The author reported that the photographic magnitude of the star vary from 14.30 to 16.60 mag. The star is included in the list of H$\alpha$ emission-line star published in Herbig (1958). The author determined the photographic magnitude of the star $m_{pg}$=15.00 mag and its spectral type K0.

Filin (1974) presented observations of V521 Cyg obtained during the period 1969--1971 in system close to $B$. The author registered slowly fluctuations in the star's brightness in the interval 14.50--15.10 mag, as these fluctuations are characterized with deep drops in the brightness of V521 Cyg from the Algol type. Chavarr\'{i}a-K. et al. (1989) observed the star in $uvby-$$\beta$ photometric system and concluded that it is PMS star.

Herbst \& Shevchenko (1999) observed V521 Cyg during the period 1986--1997 and determined the amplitude of its photometric activity $\Delta V$=1.77 mag, average stellar magnitude $\overline{V}$=13.74 mag, and colour indices $U-B$=0.58, $B-V$=1.25 and $V-R$=1.18 mag. Mitskevich \& Pavlenko (2001) presented the light curve in $V$-band of the star for the period 1999 July--October. It is seen that for about 20 days (in 1999 August), the star faded by 0.85 mag. The star spend about 5 days in minimum light state whereafter in next 70 days the star's brightness gradually returned to the previous maximum level. The authors suggest that the reason for the decline in brightness of V521 Cyg is probably obscuration from circumstellar clouds of dust. In the presented colour indices $V-I$ and $V-R$ of the star is not observed bluing effect, because the registered minimum from Mitskevich \& Pavlenko (2001) is relatively shallow.

Grankin et al. (2007) conducted long-term study of V521 Cyg during the period 1986 June--2003 September. The authors presented the light curve in $V$-band of the star with brightness variations from 13.37 to 14.70 mag. Non-linear dependence of $V-R$ colour index versus stellar $V$ magnitude is present in Grankin et al. (2007).

In $J-H$/$H-K_{s}$ diagram V521 Cyg is situated in the upper right-hand corner, close to the intrinsic T Tauri line in the CTTS location (Laugalys et al. 2006).
Poljan\v{c}i\'{c} Beljan et al. (2014) determined the period of V521 Cyg, which is found to be 503 days. The cause of the period found is likely precession of the circumstellar disk or eclipses from a second component or by clouds of gas and dust orbiting the star.


The $BVRI$ light curves of V521 Cyg from all our CCD observations (Poljan\v{c}i\'{c} Beljan et al. 2014; Ibryamov 2015a; and the present paper) are shown in Figure 3. In the figure, circles denote CCD photometric data obtained with the 2-m RCC telescope, diamonds -- the photometric data collected with the 1.3-m telescope, triangles -- the photometric data acquired with the 50/70-cm Schmidt telescope, and squares -- the photometric data taken with the 60-cm Cassegrain telescope. The results of our recent multicolour CCD observations of V521 Cyg are summarized in Table 4\footnote{The tables with photometric CCD observations and data of the stars from our study only available at the Online material of the present paper.}. The columns contains data (DD/MM/YYYY format) and Julian data (J.D.) of the observations, $IRVB$ magnitudes of the star, telescope and CCD camera used.

\begin{figure}[]
  \begin{center}
    \centering{\epsfig{file=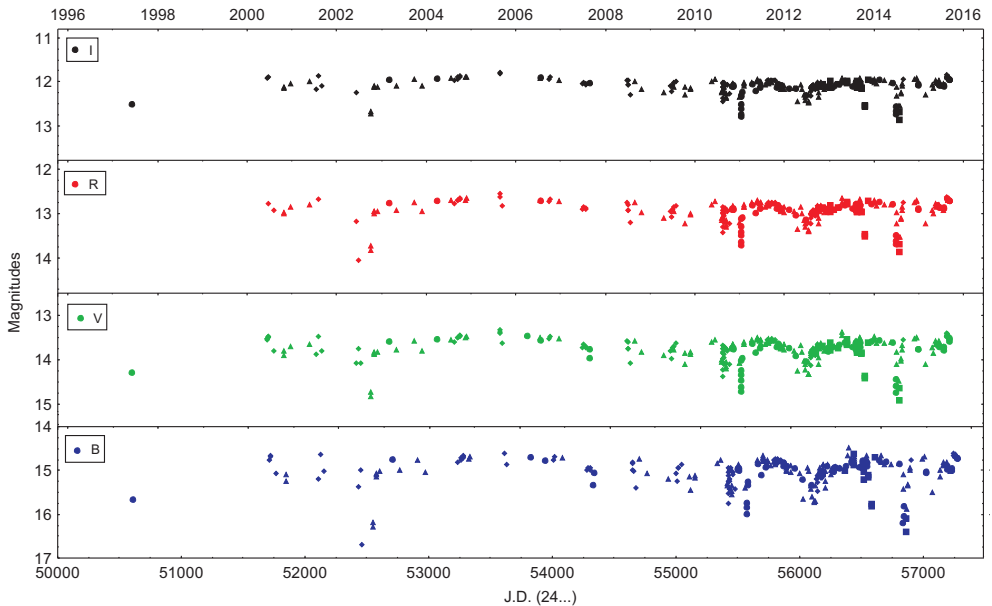, width=\textwidth}}
    \caption[]{CCD $IRVB$ light curves of V521 Cyg for the period 1997 June--2015 September}
    \label{fig3}
  \end{center}
\end{figure}

The brightness of V521 Cyg during all period of our observations 1997--2015 vary in the range 11.78--12.85 mag for $I$-band, 12.55--14.05 mag for $R$-band, 13.32--14.89 mag for $V$-band, and 14.49--16.69 mag in $B$-band. The observed amplitudes are $\Delta I$=1.07 mag, $\Delta R$=1.50 mag, $\Delta V$=1.57 mag, and $\Delta B$=2.20 mag.

It can be seen from Fig. 3 that the star spends most of the time at high light. During our photometric monitoring, several fading events of V521 Cyg are registered. Ibryamov et al. (2015a) described seven deep declines in the brightness observed in all bands. The drops in the brightness are non-periodic and have different duration and amplitudes. It is possible to suggest the existence of a frequent deep fading events during periods with insufficient data.
The new data shows one fading event observed in the beginning of 2015 with amplitudes 0.40 mag in $I$-band, 0.48 mag in $R$-band, 0.57 mag in $V$-band, and 0.81 mag for $B$-band.

The results obtained during our long-term study support the hypothesis of variable extinction as a reason for variability of V521 Cyg (Mitskevich \& Pavlenko (2001) and Grankin et al. (2007)). The large amplitudes of observed declines in the brightness of the star (Fig. 3) are indication of UXor-type variability and the deep fading events presumably are result of circumstellar dust or clouds obscuration.

The measured colour indices $V-R$, $V-R$, and $B-V$ versus the stellar $V$ magnitude during the period of our observations are plotted in Fig. 4. Nevertheless of the large amplitude of variability, the substantial bluing effect is not seen on the figure. 
A possible explanation of this phenomenon is that the declines in the brightness at different periods are caused by clouds of dust and particles of various sizes.
It can be supposed that in this case we observe eclipses from a combination of clouds with different structure rotating in orbits around the star.

\begin{figure}
\begin{center}
\includegraphics[width=4cm]{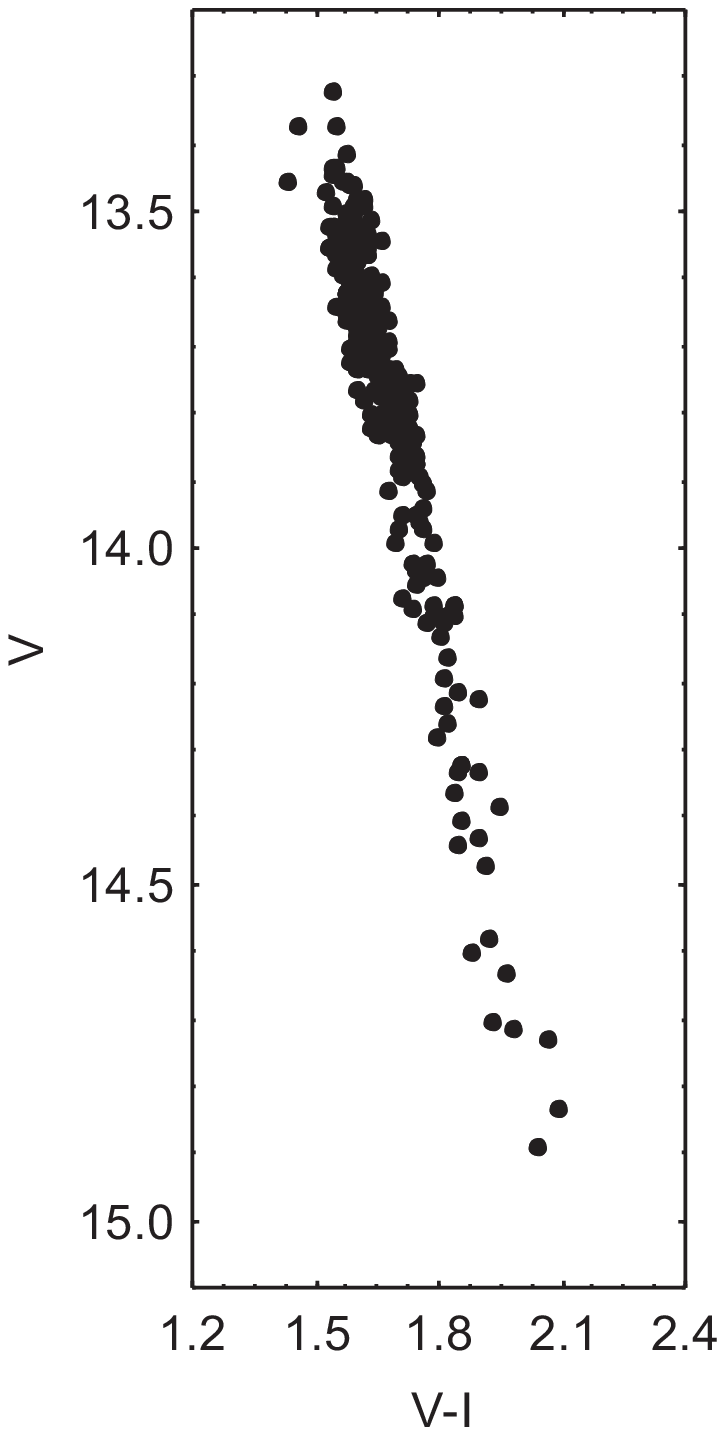}
\includegraphics[width=4cm]{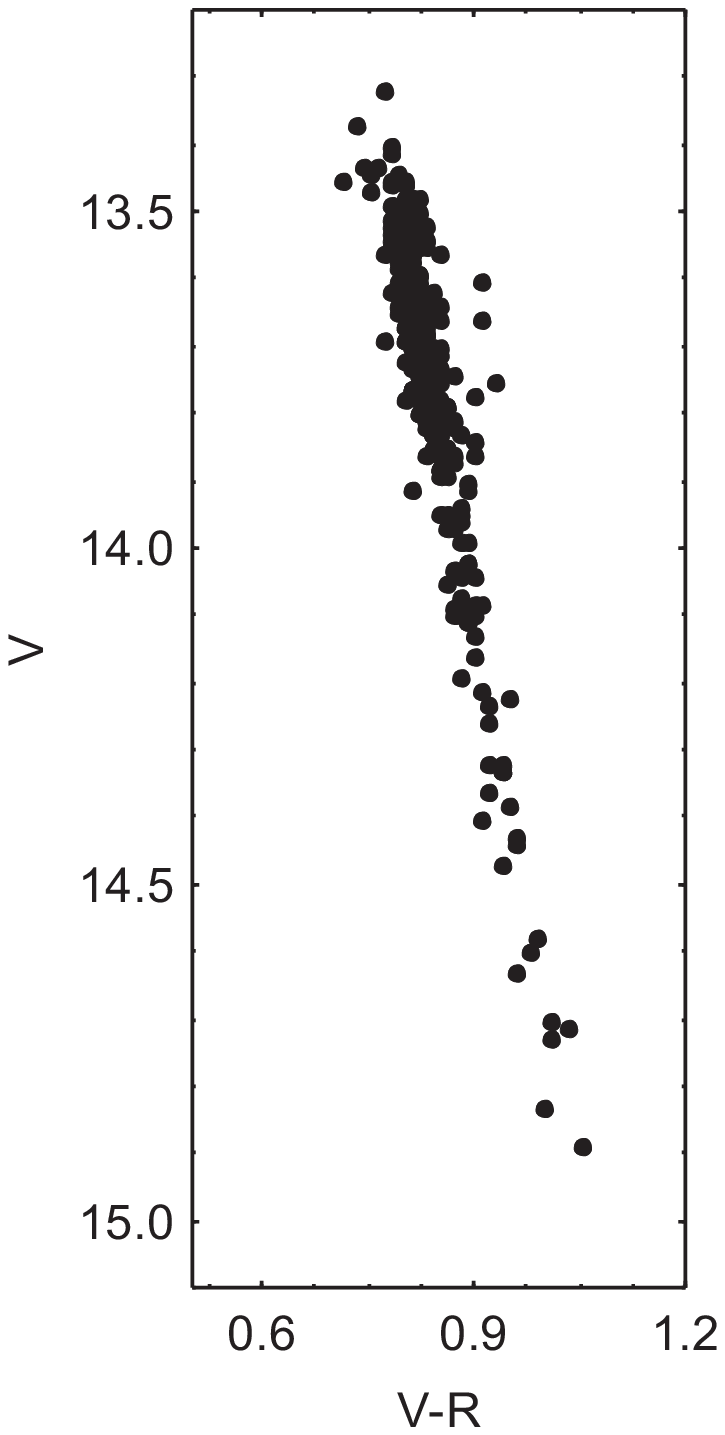}
\includegraphics[width=4cm]{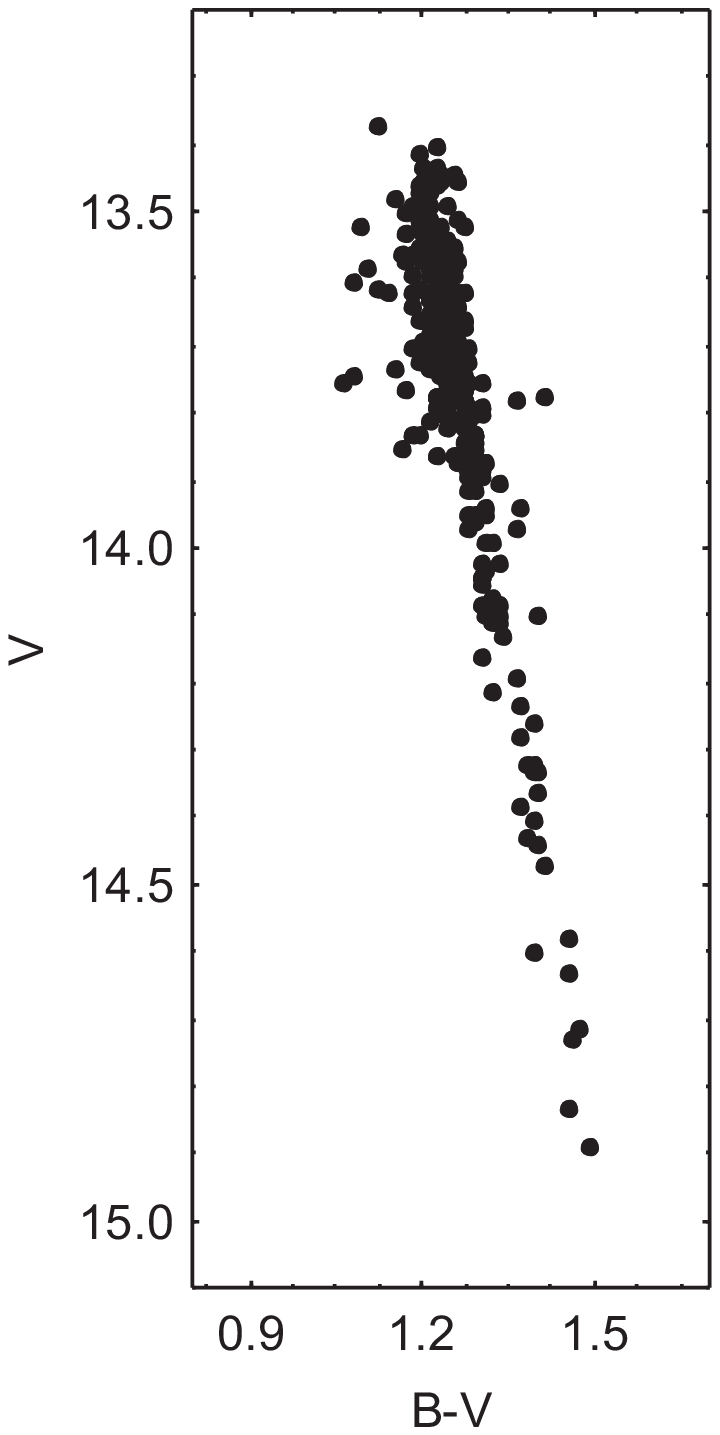}
\caption{Relationship between $V$ magnitude and the $V-I$, $V-R$, and $B-V$ colour indices of V521 Cyg in the period 1997$-$2015}\label{fig4}
\end{center}
\end{figure}

The $BVRI$ light curves of V521 Cyg look very similar to light curves of FHO 27, which was classified as UXor by Ibryamov (2015b). FHO 27 is located at about 9$\arcmin$ from V521 Cyg.


\subsection*{3.2. V752 Cyg} 

The variability of V752 Cyg was reported in Erastova \& Tsvetkov (1978). Kohoutek \& Wehmeyer (1997) confirmed the variability of the star, measured its photographic magnitude $m_{pg}$=15.30 mag and registered H$\alpha$ emission line in its spectrum. V752 Cyg is included in the list of YSO candidates published in Guieu et al. (2009).


The $BVRI$ light curves of V752 Cyg from all our CCD observations (Poljan\v{c}i\'{c} Beljan et al. 2014; Ibryamov 2015a; and the present paper) are shown in Fig. 5. The symbols used for the different telescopes are as in Fig. 3. The results from our recent multicolour CCD observations of V752 Cyg are summarized in Table 5. The columns have the same contents as in Table 4.

\begin{figure}[]
  \begin{center}
    \centering{\epsfig{file=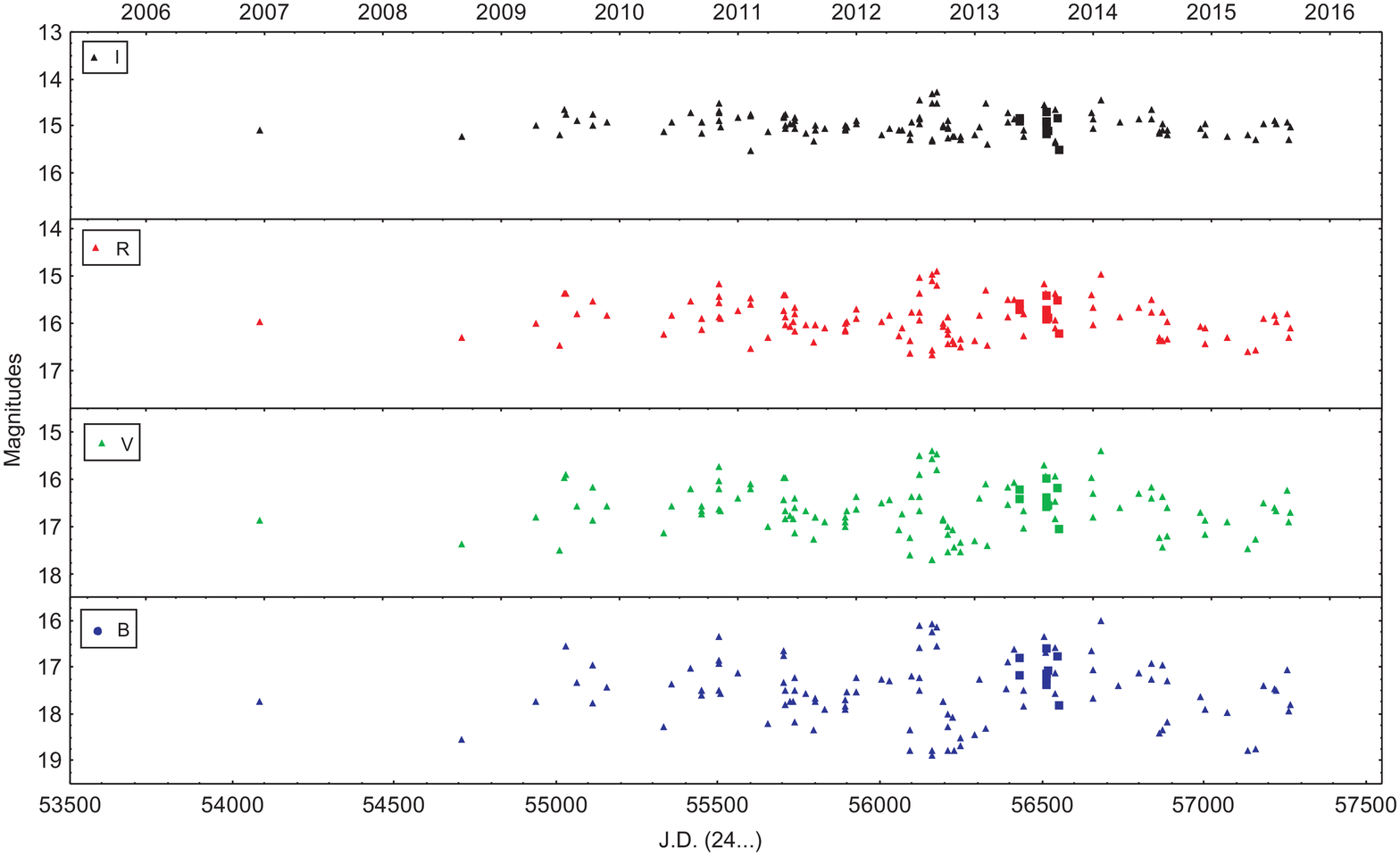, width=\textwidth}}
    \caption[]{CCD $IRVB$ light curves of V752 Cyg for the period 2006 December--2015 September}
    \label{fig5}
  \end{center}
\end{figure}

It can be seen from Fig. 5 that during our study the brightness of V752 Cyg vary around some intermediate level. The star shows principally increases in the brightness with different amplitudes.
The brightness of the star during the period of our observations 2006--2015 vary in the range 14.28--15.55 mag for $I$-band, 14.89--16.67 mag for $R$-band, 15.38--17.71 mag for $V$-band, and 16.02--18.89 mag in $B$-band. The observed amplitudes are $\Delta I$=1.27 mag, $\Delta R$=1.78 mag, $\Delta V$=2.33 mag, and $\Delta B$=2.87 mag.

The colour indices $V-I$, $V-R$, and $V-B$ versus $V$ magnitude during the period of our observations are plotted in Fig. 6. From the figure, it is seen that the star becomes bluer when it increases its brightness. Such colour variations are indication for flare events. These flares can be explained by increase in accretion rate from the circumstellar disk onto the stellar surface. Evidences of periodicity in the brightness variability are not detected.
In the $J-H / H-K_{s}$ diagram (Fig. 2) V752 Cyg lies by 0.05 mag above the intrinsic T Tauri line -- it have negligible infrared excess. 

\begin{figure}
\begin{center}
\includegraphics[width=4cm]{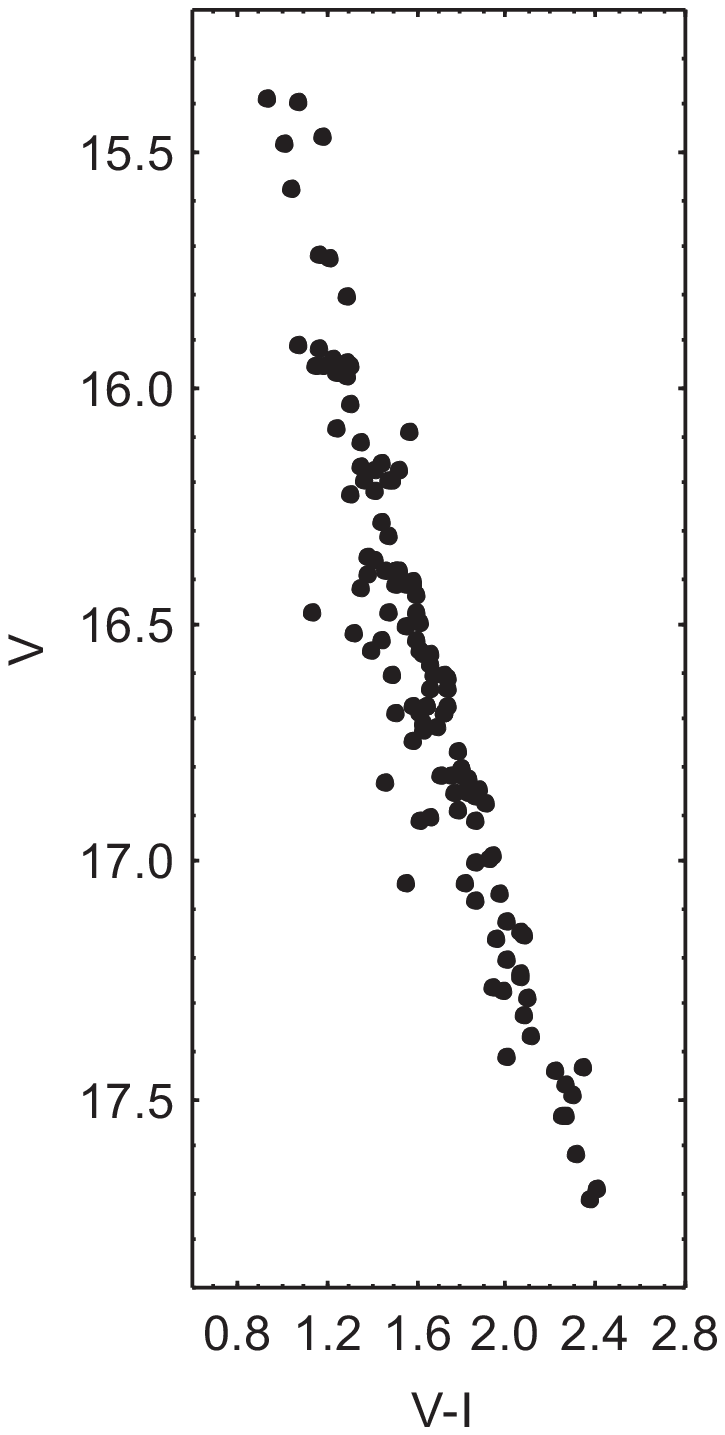}
\includegraphics[width=4cm]{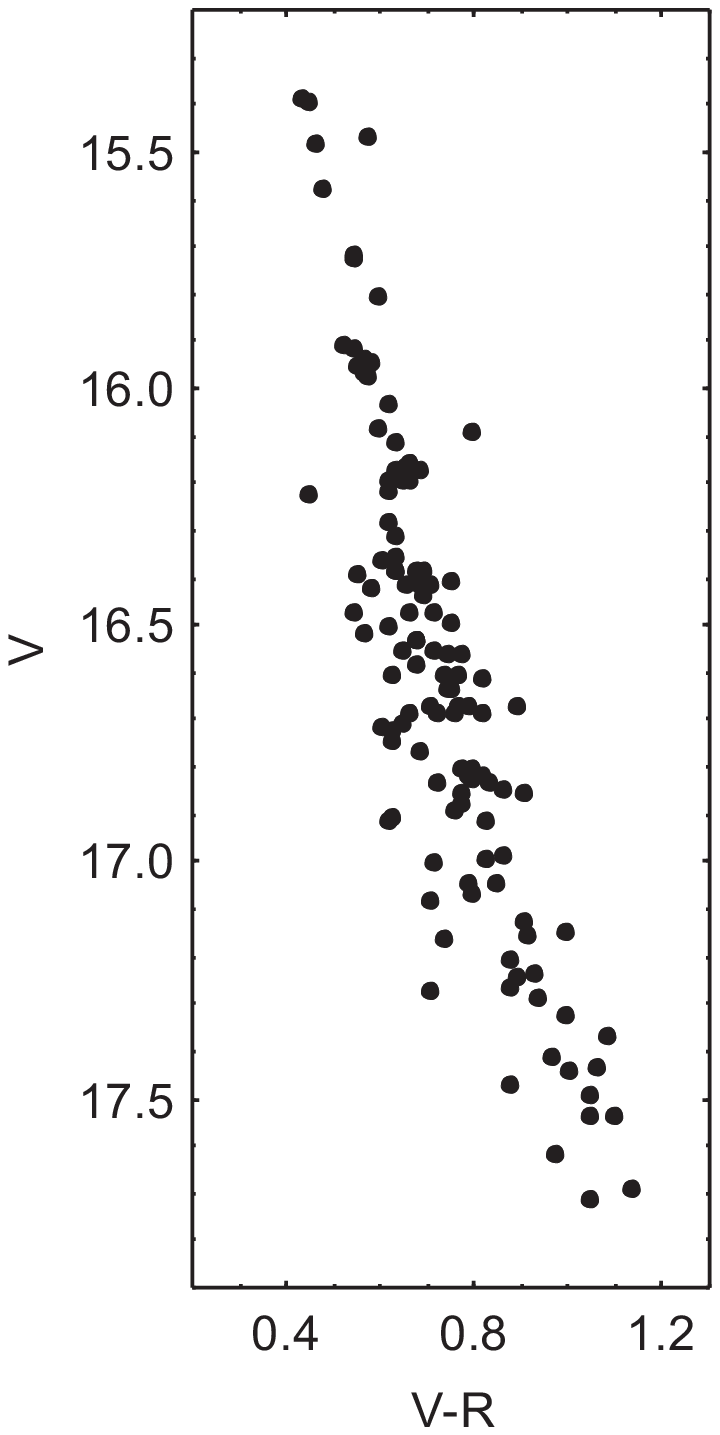}
\includegraphics[width=4cm]{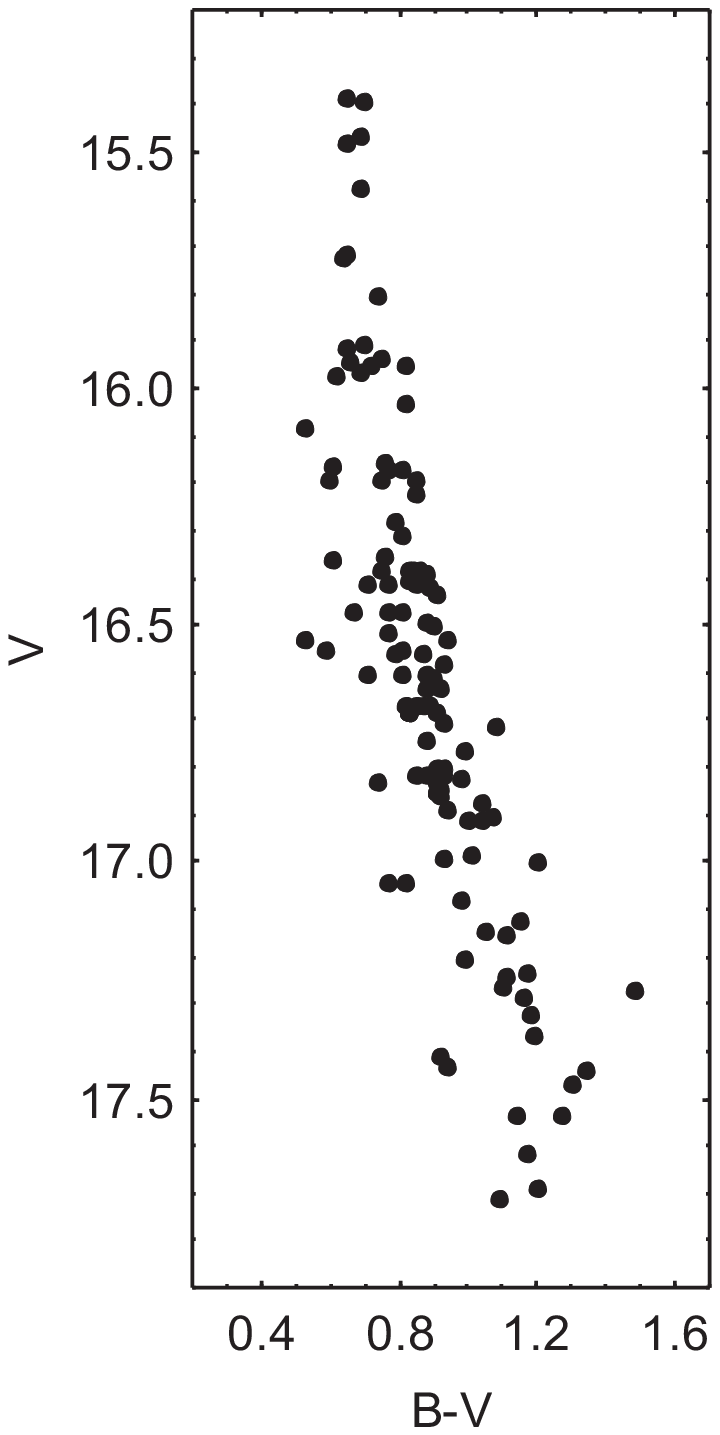}
\caption{Relationship between $V$ magnitude and the $V-I$, $V-R$, and $B-V$ colour indices of V752 Cyg in the period 2006$-$2015}\label{fig6}
\end{center}
\end{figure}


\subsection*{3.3. V1538 Cyg} 

The star V1538 Cyg was discovered and classified as a flare star by Erastova \& Tsvetkov (1974). The authors reported a flare event on 1973 July 30 and determined its photographic magnitude $m_{pg}$=18.50 mag and $\Delta U$=2.50 mag. 
Laugalys et al. (2006) registered H$\alpha$ emission-line in the star's spectrum and measured its magnitude $V$=17.01 mag and colour index $U-V$=4.32 mag. The authors determined the spectral type as M2 and the distance to the star as $r$=248 pc.

The spectrum of V1538 Cyg taken on 2007 October 23 by Corbally et al. (2009) shows a M1 photosphere with low emission in H$\alpha$. In $J-H$/$H-K_{s}$ diagram the star lies in the upper part of the early M dwarf band, so there are no indications of the presence of the envelope. This is confirmed by the $Spitzer$ fluxes -- the energy distribution of the star no showed infrared radiation. According to Corbally et al. (2009) V1538 Cyg can be either an unreddened field dwarf with chromospheric activity, located at 350 pc, or a post-T Tauri star at the front edge of the NGC 7000/IC 5070 complex. Armond et al. (2011) included V1538 Cyg in the list of H$\alpha$ emission-line star and measured its magnitudes as $I$=16.28, $R$=16.85 and $V$=17.82 mag.


The multicolour $BVRI$ light curves of V1538 Cyg from all our CCD observations in the period 2000--2015 (Poljan\v{c}i\'{c} Beljan et al. 2014; and the present paper) are shown in Fig. 7. The symbols used for the different telescopes are as in Fig. 3. The results of our recent multicolour CCD observations of the star are summarized in Table 6. The columns have the same contents as in Table 4.

\begin{figure}[]
  \begin{center}
    \centering{\epsfig{file=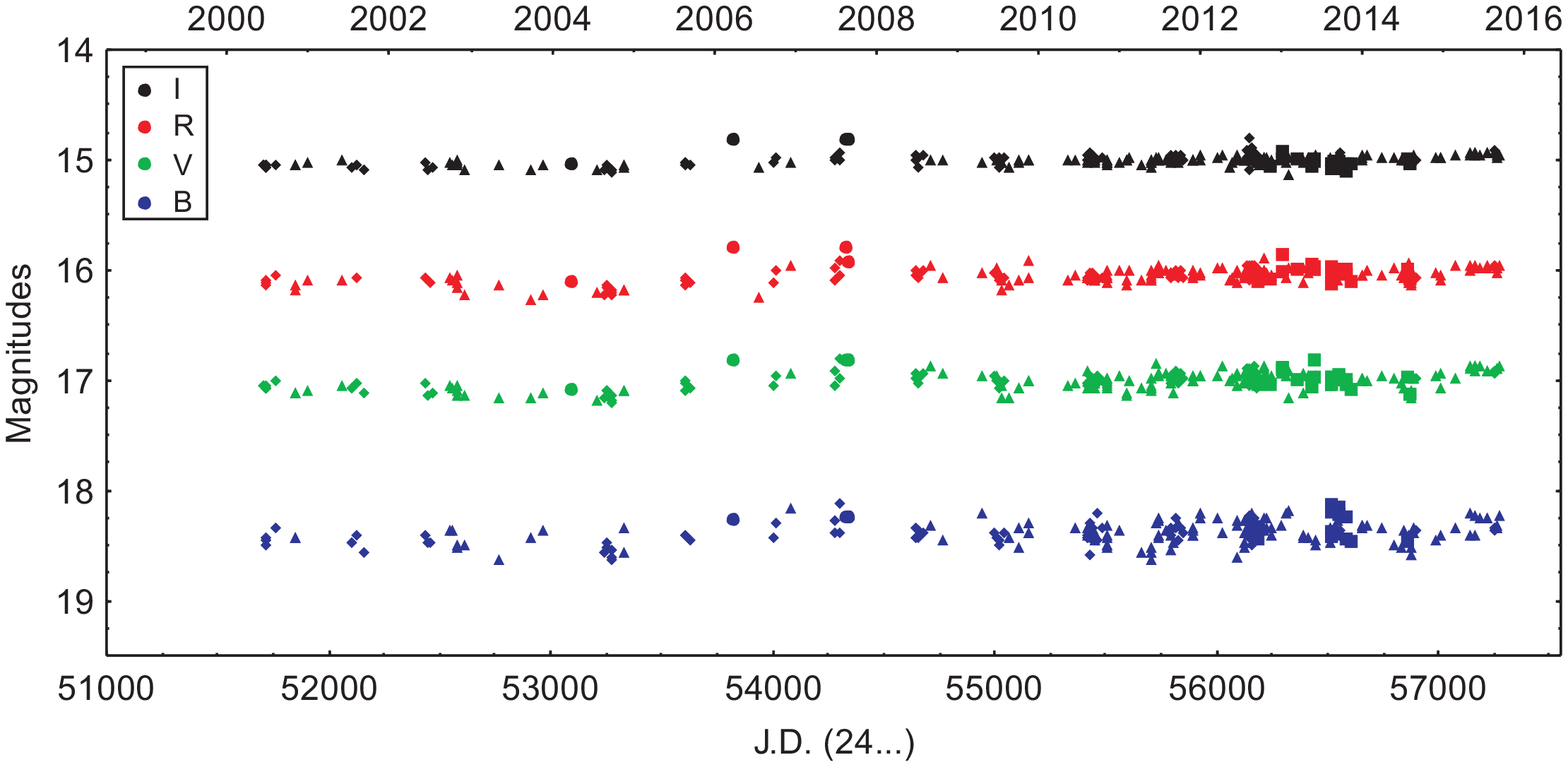, width=\textwidth}}
    \caption[]{CCD $IRVB$ light curves of V1538 Cyg for the period 2000 June--2015 September}
    \label{fig7}
  \end{center}
\end{figure}

From Fig. 7 can be seen that during the period of our observations V1538 Cyg shows irregular variability in all bands, but significant flare events are not registered.
The brightness of the star during the period 2000--2015 vary in the range 14.79--15.13 mag for $I$-band, 15.78--16.26 mag for $R$-band, 16.80--17.21 mag for $V$-band, and 18.11--18.64 mag for $B$-band. The observed amplitudes are $\Delta I$=0.34 mag, $\Delta R$=0.48 mag, $\Delta V$=0.41 mag, and $\Delta B$=0.53 mag in the same period. 

Usually, such low amplitude variability is typical for low-mass WTTS, whose variability is produced by rotation of the spotted surface. This is confirmed by the position of the star in the 2MASS two-colour diagram in Fig. 2 -- there are no indications of the presence of infrared excess.
Evidences of periodicity in the brightness variability of V1538 Cyg are not detected.


\subsection*{3.4. V1539 Cyg} 

The star V1539 Cyg was discovered and classified as a H$\alpha$ emission-line star by Herbig (1958) with photographic magnitude $m_{pg}$=17.00 mag. Welin (1973) confirmed it as a H$\alpha$ emission star with strong H$\alpha$ intensity and determined $V$=15.50 and $B$=17.00 mag. 

Gieseking \& Schumann (1976) discovered that star's brightness vary with amplitude greater than 0.25 mag, which led to the conclusion of the suspected variability. Herbig \& Bell (1988) classified V1539 Cyg as a T Tauri star with $m_{pg}$=16.30 mag. Kohoutek \& Wehmeyer (1997) confirmed the variability of the star with maximum brightness in photoelectric $V$ system 14.50 mag. and determined its spectral type as K6IV.

The spectrum of V1539 Cyg taken on 2007 October 21 by Corbally et al. (2009) shows a G5e photosphere with strong emission in H$\alpha$. CaII and OI lines are also registered in emission. In $J-H$/$H-K_{s}$ diagram the star lies by 0.12 mag above the intrinsic T Tauri line. The $Spitzer$ fluxes exhibit the presence of a considerable thermal emission from the dust envelope (Corbally et el. 2009).
According to Findeisen et al. (2013) V1539 Cyg shows bursts events and they reported average stellar magnitude $\overline{R}$=14.60 mag and $\Delta R$=0.70 mag.


The $BVRI$ light curves of V1539 Cyg from all our CCD observations (Poljan\v{c}i\'{c} Beljan et al. 2014; Ibryamov 2015a; and the present paper) are shown in Fig. 8. The symbols used for the different telescopes are as in Fig. 3. The results of our recent multicolour CCD observations of the star are summarized in Table 7. The columns have the same contents as in Table 4.

\begin{figure}[]
  \begin{center}
    \centering{\epsfig{file=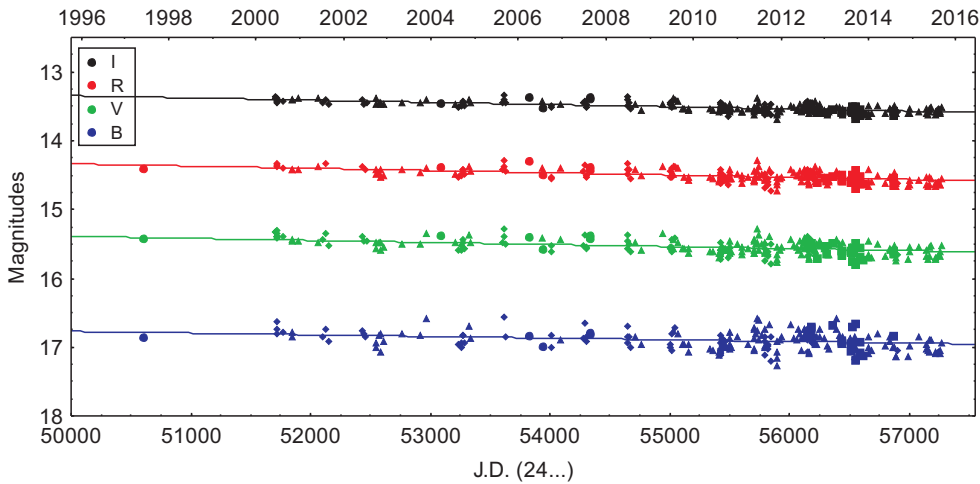, width=\textwidth}}
    \caption[]{CCD $IRVB$ light curves of V1539 Cyg for the period 1997 June--2015 September}
    \label{fig8}
  \end{center}
\end{figure}

The brightness of V1539 Cyg during the period of our observations 1997--2015 vary in the range 13.34--13.68 mag for $I$-band, 14.28--14.72 mag for $R$-band, 15.28--15.79 mag for $V$-band, and 16.56--17.26 mag for $B$-band. The observed amplitudes are $\Delta I$=0.34 mag, $\Delta R$=0.44 mag, $\Delta V$=0.51 mag, and $\Delta B$=0.70 mag. 
From Fig. 8 it can be seen that during the period of our observations V1539 Cyg shows variability in all bands. This variability include short rises and decreases of the star's brightness with small amplitudes.

Important result from our long-term photometric study of V1539 Cyg is that during the whole period of observations the total star's brightness gradually decreases. Using a linear approximation for all our data of the star, we calculated the following values for the rates of decreases: $~$0.0117 mag $yr^{-1}$ for $I$-band, $~$0.0119 mag $yr^{-1}$ for $R$-band, $~$0.0109 mag $yr^{-1}$ for $V$-band, and $~$0.0090 mag $yr^{-1}$ for $B$-band.

The light curves of the star gives grounds to predict different reasons for observed variability of the star -- existence of hot and cool spots on the stellar surface, and/or irregular obscuration of the star by circumstellar material. Evidences of periodicity in the brightness variability are not detected.


\subsection*{3.5. V1716 Cyg} 

The variability of V1716 Cyg was discovered by Erastova \& Tsvetkov (1978). The authors reported irregular brightness fluctuations and measured its photographic magnitudes $m_{pg}$=17.00--17.50 mag (limit) and $U$=16.10--17.50 mag (limit).
Armond et al. (2011) included V1716 Cyg in the list of H$\alpha$ emission-line star and measured its magnitudes as $I$=15.95, $R$=16.74 and $V$=17.93 mag. Findeisen et al. (2013) discovered two bursts events of the star, which are separated by 35 days, as the first one lasting 5--20 days, and the second one -- 3 days. The authors reported average stellar magnitude $\overline{R}$=16.50 mag and $\Delta R$=1.10 mag.

Poljan\v{c}i\'{c} Beljan et al. (2014) determined the period of V1716 Cyg, which is found to be 4.15 days. This periodicity is probably connected with the stellar rotation and the position of the dark spots on the stellar surface.


The $BVRI$ light curves of V1716 Cyg from all our CCD observations (Poljan\v{c}i\'{c} Beljan et al. 2014; Ibryamov 2015a; and the present paper) are shown in Fig. 9. The symbols used for the different telescopes are as in Fig. 3. The results of our recent multicolour CCD observations of the star are summarized in Table 8. The columns have the same contents as in Table 4.

\begin{figure}[]
  \begin{center}
    \centering{\epsfig{file=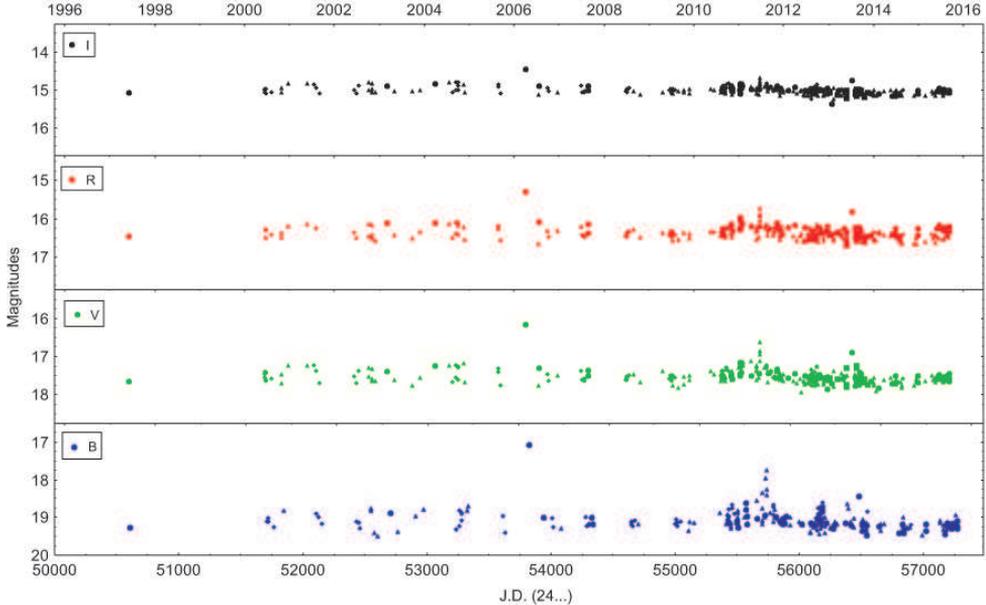, width=\textwidth}}
    \caption[]{CCD $IRVB$ light curves of V1716 Cyg for the period 1997 June--2015 September}
    \label{fig9}
  \end{center}
\end{figure}

The brightness of V1716 Cyg during the period of our observations 1997--2015 vary in the range 14.43--15.36 mag for $I$-band, 15.28--16.68 mag for $R$-band, 16.14--17.93 mag for $V$-band, and 17.05--19.50 mag for $B$-band. The observed amplitudes are $\Delta I$=0.93 mag, $\Delta R$=1.40 mag, $\Delta V$=1.79 mag, and $\Delta B$=2.45 mag in the same period. 

Ibryamov et al. (2015a) described four eruptive events of V1716 Cyg observed in all bands. The irregular flares can be explained with short-lived accretion-related events at the stellar surface or as flares from UV Cet-type. Other irregular variations of the star's brightness, observed in Fig. 9 are with smaller amplitudes and are caused probably by rotating hot and cool spots on the stellar surface. These results are an indication that V1716 Cyg is likely CTTS.

In the $J-H$/$H-K_{s}$ diagram (Fig. 2) the star lies by 0.11 mag above the intrinsic T Tauri line, i.e., it have infrared excess and this is indication of the presence of the envelope.
Our new data confirm the periodicity of V1716 Cyg found by Poljan\v{c}i\'{c} Beljan et al. (2014).


\subsection*{3.6. V1957 Cyg} 

V1957 Cyg was discovered and classified as a flare star by Chavushian \& Jankovics (1985). The authors reported a flare event on 1979 October 19 when the brightness of the star in $U$-band increased from 18.00 to 16.00 mag. In the following series of observations performed in 1981 and 1984 flare events are not registered.
Laugalys et al. (2006) classified V1957 Cyg as a K6 dwarf with possible H$\alpha$ emission. The authors measured $V$=16.30 mag and colour index $U-V$=4.05 mag, and determined the distance to the star as $r$=481 pc.

The spectrum of V1957 Cyg taken on 2007 October 22 by Corbally et al. (2009) shows a Ge photosphere with weak emission in H$\alpha$. The authors reported that in $J-H$/$H-K_{s}$ diagram the star lies close to the intrinsic position of M0V star, but it can be a reddened G-type star, too. IRAC magnitudes measured by $Spitzer$ show no infrared excess. According to authors V1957 Cyg may have been in a particularly active phase at the time of photometric observations, and in quiet stage during the spectral observations. 


The $BVRI$ light curves of the star from all our CCD observations (Poljan\v{c}i\'{c} Beljan et al. 2014; and the present paper) are shown in Fig. 10. The symbols used for the different telescopes are as in Fig. 3. The results of our recent multicolour CCD observations of the star are summarized in Table 9. The columns have the same contents as in Table 4.

\begin{figure}[]
  \begin{center}
    \centering{\epsfig{file=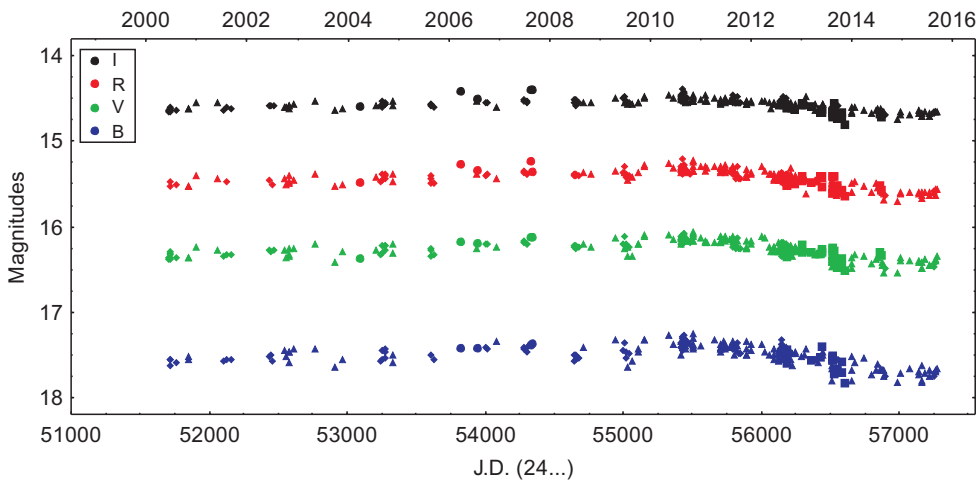, width=\textwidth}}
    \caption[]{CCD $IRVB$ light curves of V1957 Cyg for the period 2000 June--2015 September}
    \label{fig10}
  \end{center}
\end{figure}

The brightness of V1957 Cyg during the period 2000--2015 vary in the range 14.38--14.79 mag. for $I$-band, 15.20--15.70 mag. for $R$-band, 16.06--16.54 mag. for $V$-band, and 17.25--17.83 mag. for $B$-band. The observed amplitudes are $\Delta I$=0.41 mag, $\Delta R$=0.50 mag, $\Delta V$=0.48 mag, and $\Delta B$=0.58 mag during the same period.

It can be seen from Fig. 10 that the star's brightness increasing smoothly to the mid-2010, then it begins decrease smoothly to the mid-2014. From the mid-2014 so far the star's brightness vary around some intermediate level with very small amplitudes. From the observed amplitudes of V1957 Cyg and its position in the two-colour diagram (Figure 2) it can be assumed that the star is probably M dwarf with chromospheric activity.
Evidences of periodicity in the brightness variability of the star are not detected.


\subsection*{3.7. V2051 Cyg}

V2051 Cyg was discovered and classified as a flare star by Parsamian et al. (1994). The authors reported a flare event on 1977 September 07 when the brightness of the star in $U$-band increased from the plate limit to 14.00 mag with amplitude $\Delta m_{U}$>4.00 mag.

Laugalys et al. (2006) classified V2051 Cyg as a possible T Tauri star with $V$=16.59 mag. The spectrum of the star taken on 2007 October 21 by Corbally et al. (2009) shows a M3.5e photosphere with weak emission in H$\alpha$. In $J-H$/$H-K_{s}$ diagram the star lies close to the intrinsic position of M3V stars with no infrared excess. Its energy distribution, constructed from the 2MASS and $Spitzer$ observations, also suggest for a normal star without envelope. 
According to Corbally et al. (2009) V2051 Cyg is most probably a M dwarf with weak chromospheric activity and the distance to the star is $r$=137 pc, i.e., the star cannot have any relation to the NGC 7000/IC 5070 complex.

Poljan\v{c}i\'{c} Beljan et al. (2014) determined the period of the star, which is found to be 384 days, likely caused by precession of the circumstellar disk or by clouds of gas and dust orbiting the star.


The multicolour $BVRI$ light curves of V2051 Cyg from all our CCD observations in the period 2000--2015 (Poljan\v{c}i\'{c} Beljan et al. 2014; Ibryamov 2015a; and the present paper) are shown in Fig 11. The symbols used for the different telescopes are as in Fig 3. The results of our recent multicolour CCD observations of the star are summarized in Table 10. The columns have the same contents as in Table 4.

\begin{figure}[]
  \begin{center}
    \centering{\epsfig{file=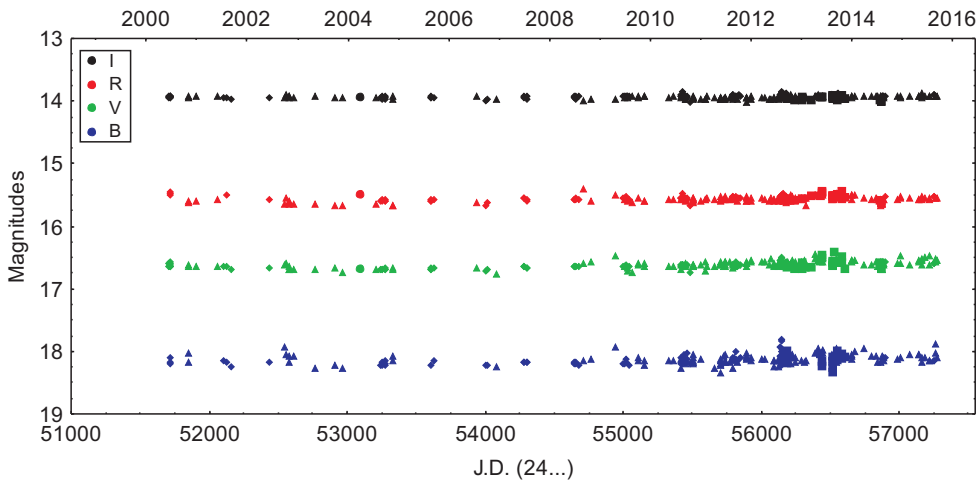, width=\textwidth}}
    \caption[]{CCD $IRVB$ light curves of V2051 Cyg for the period 2000 June--2015 September}
    \label{fig11}
  \end{center}
\end{figure}

During the period of our observations flare events are not registered, except a few low amplitude increases in brightness in $V$- and $B$-band.
The brightness of the star during the period 2000--2015 vary in the range 13.84--14.02 mag for $I$-band, 15.40--15.68 mag for $R$-band, 16.40--16.77 mag for $V$-band, and 17.80--18.35 mag for $B$-band. The observed amplitudes are $\Delta I$=0.18 mag, $\Delta R$=0.28 mag, $\Delta V$=0.37 mag, and $\Delta B$=0.55 mag in the same period. 
Low amplitude of variability and relatively stable light variations confirm the assumptions on the nature of V2051 Cyg as a M dwarf with weak chromospheric activity (Corbally et al. 2009).


\subsection*{3.8. LkH$\alpha$ 186}

The star LkH$\alpha$ 186 was discovered and classified as a H$\alpha$ emission-line star by Herbig (1958) with photographic magnitude $m_{pg}$=18.00 mag. Herbig \& Bell (1988) classified it as a T Tauri variable with spectral type K5. Cohen \& Kuhi (1979) measured $V$=18.00 mag of the star. In the paper of Weintraub (1990) LkH$\alpha$ 186 is also classified as a T Tauri variable.

Guieu et al. (2009) included the star in their list of YSO candidates and measured $I$=14.747, $V$=17.395, and $B$=19.322 mag. Armond et al. (2011) included it in the list of H$\alpha$ emission-line star and measured $I$=15.85, $R$=16.68 and $V$=18.12 mag and classified it as a CTTS.


The $BVRI$ light curves of LkH$\alpha$ 186 from all our CCD observations (Poljan\v{c}i\'{c} Beljan et al. 2014; and the present paper) are shown in Fig. 12. The symbols used for the different telescopes are as in Fig. 3. The results of our recent multicolour CCD observations of the star are summarized in Table 11. The columns have the same contents as in Table 4.

\begin{figure}[]
  \begin{center}
    \centering{\epsfig{file=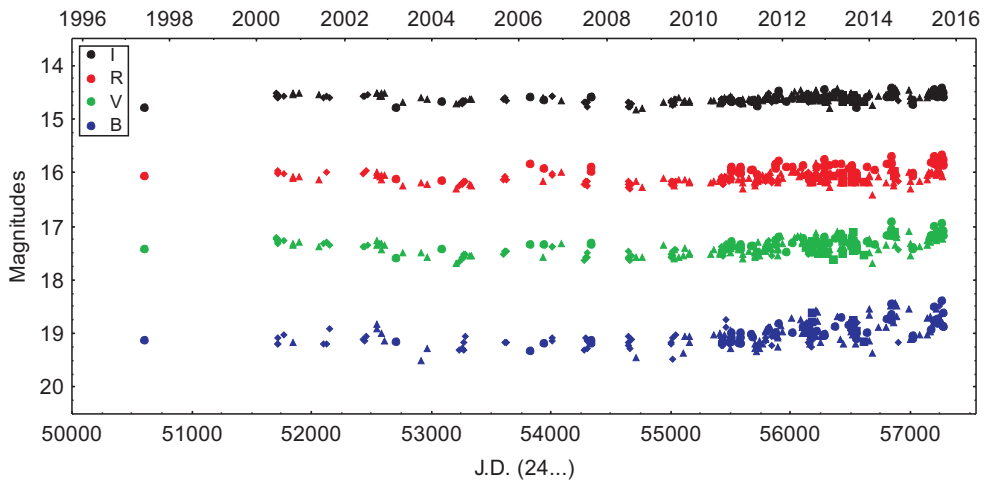, width=\textwidth}}
    \caption[]{CCD $IRVB$ light curves of LkH$\alpha$ 186 for the period 1997 June--2015 September}
    \label{fig12}
  \end{center}
\end{figure}

The brightness of the star during the period 1997--2015 vary in the range 14.39--14.82 mag for $I$-band, 15.65--16.42 mag for $R$-band, 16.91--17.68 mag for $V$-band, and 18.37--19.50 mag for $B$-band. The observed amplitudes are $\Delta I$=0.43 mag, $\Delta R$=0.77 mag, $\Delta V$=0.77 mag, and $\Delta B$=1.13 mag.

It can be seen from Fig. 12 that the star usually spends most of the time at low light. LkH$\alpha$ 186 exhibit increases of the brightness with different amplitudes probably due to the presence of hot spots on the stellar surface. The variability is produced by variable accretion from the circumstellar disk and it is typical of CTTS. 

In the two-colour diagram (Fig. 2) the star lies by 0.37 mag above the intrinsic T Tauri line. The star have clear infrared excess, indicating the presence of a circumstellar disk. Evidences of periodicity in the brightness variability are not detected.


\subsection*{3.9. LkH$\alpha$ 187}

The star LkH$\alpha$ 187 was discovered and classified as a H$\alpha$ emission-line star by Herbig (1958) with photographic magnitude $m_{pg}$=18.50. Herbig \& Bell (1988) classified it as a T Tauri variable with spectral type K3. Cohen \& Kuhi (1979) measured $V$=18.10 mag of the star. In the work of Weintraub (1990) LkH$\alpha$ 187 is also classified as a T Tauri variable.

Guieu et al. (2009) included the star in their list of YSO candidates and measured $I$=14.811, $V$=17.639, and $B$=19.476 mag. Armond et al. (2011) included it in the list of H$\alpha$ emission-line star and measured $I$=16.00, $R$=17.05 and $V$=18.56 mag and classified it as a CTTS.


The $BVRI$ light curves of LkH$\alpha$ 187 from all our CCD observations (Poljan\v{c}i\'{c} Beljan et al. 2014; and the present paper) are shown in Fig. 13. The symbols used for the different telescopes are as in Fig. 3. The results of our recent multicolour CCD observations of the star are summarized in Table 12. The columns have the same contents as in Table 4.

\begin{figure}[]
  \begin{center}
    \centering{\epsfig{file=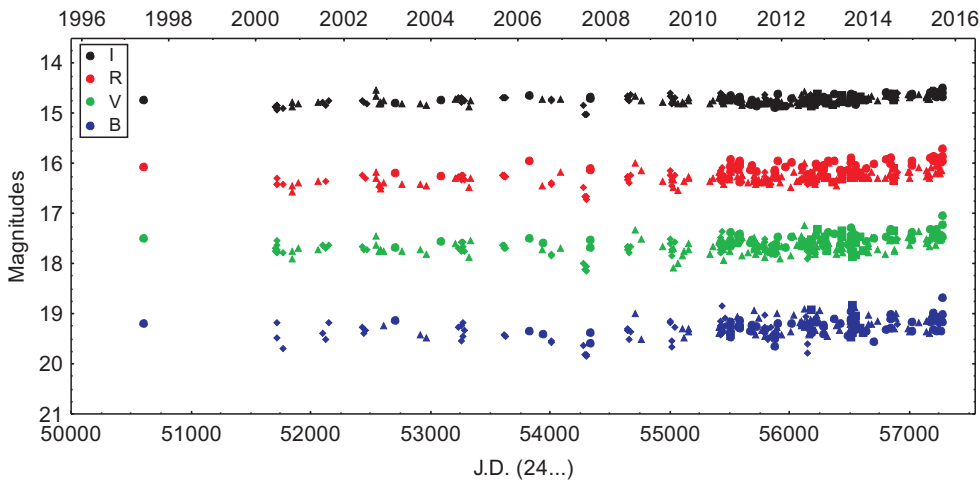, width=\textwidth}}
    \caption[]{CCD $IRVB$ light curves of LkH$\alpha$ 187 for the period 1997 June--2015 September}
    \label{fig13}
  \end{center}
\end{figure}

The brightness of the star during the period 1997--2015 vary in the range 14.46--15.02 mag for $I$-band, 15.68--16.71 mag for $R$-band, 17.02--18.15 mag for $V$-band, and 18.66--19.85 mag for $B$-band. The observed amplitudes are $\Delta I$=0.56 mag, $\Delta R$=1.03 mag, $\Delta V$=1.13 mag, and $\Delta B$=1.19 mag.

It can be seen from Fig. 13 that similar to LkH$\alpha$ 186, LkH$\alpha$ 187 usually spends most of the time at low light. The star exhibit increases of the brightness with different amplitudes. This variability is due likely to the presence of hot spots on the stellar surface, i.e., it is produced by strong variable accretion from the circumstellar disk.

LkH$\alpha$ 187 shows variability typical of CTTS. In the 2MASS two-colour diagram (Fig. 2) the star lies by 0.38 mag above the intrinsic T Tauri line. LkH$\alpha$ 187 have clear infrared excess, indicating the presence of a circumstellar disk. Evidences of periodicity in the brightness variability are not detected.


\subsection*{3.10. LkH$\alpha$ 189}

The star LkH$\alpha$ 189 was discovered and classified as a H$\alpha$ emission-line star by Herbig (1958) with photographic magnitude $m_{pg}$=17.50. Welin (1973) confirmed it as an H$\alpha$ emission-line star. Herbig \& Bell (1988) classified it as a T Tauri variable with spectral type K6 and photographic magnitude 17.00 mag. Cohen \& Kuhi (1979) measured $V$=17.00 mag of the star. In the paper of Weintraub (1990) LkH$\alpha$ 189 is also  classified as a T Tauri variable.

Laugalys et al. (2006) classified LkH$\alpha$ 189 as a T Tauri star with $V$=16.69 mag and photographic spectral type K6. The position of the star in the two-colour $J-H$/$H-K_{s}$ diagram obtained by the Laugalys et al. (2006) and Meyer et al. (1997) suggest increased disk thickness and interstellar reddening. Guieu et al. (2009) included the star in their list of YSO candidates and measured $I$=13.936, $V$=16.376, and $B$=18.175 mag. Armond et al. (2011) included it in the list of H$\alpha$ emission-line star and measured $I$=15.03, $R$=15.80 and $V$=17.10 mag and classified it as a CTTS.


The $BVRI$ light curves of LkH$\alpha$ 189 from all our CCD observations (Poljan\v{c}i\'{c} Beljan et al. 2014; and the present paper) are shown in Fig. 14. The symbols used for the different telescopes are as in Fig. 3. The results of our recent multicolour CCD observations of the star are summarized in Table 13. The columns have the same contents as in Table 4.

\begin{figure}[]
  \begin{center}
    \centering{\epsfig{file=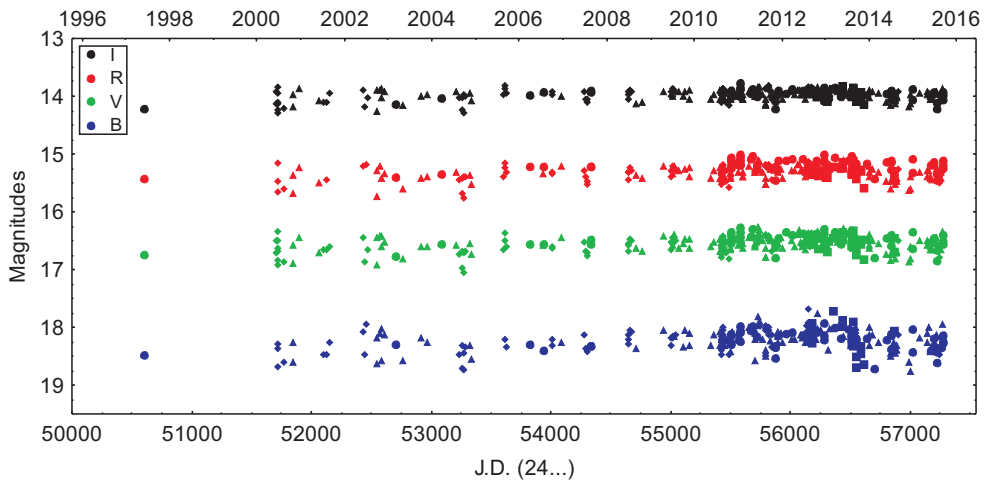, width=\textwidth}}
    \caption[]{CCD $IRVB$ light curves of LkH$\alpha$ 189 for the period 1997 June--2015 September}
    \label{fig14}
  \end{center}
\end{figure}

The brightness of LkH$\alpha$ 189 during the period 1997--2015 vary in the range 13.76--14.30 mag for $I$-band, 14.99--15.76 mag for $R$-band, 16.25--17.06 mag for $V$-band, and 17.69--18.75 mag for $B$-band. The observed amplitudes are $\Delta I$=0.54 mag, $\Delta R$=0.77 mag, $\Delta V$=0.81 mag, and $\Delta B$=1.06 mag in the same period.

During our study the brightness of LkH$\alpha$ 189 vary around some intermediate level (Fig. 14). Both increases and drops in the brightness with different amplitudes are observed. The observed drops of the star's brightness likely are caused by the existence of cool spots or groups of spots on the stellar surface or by irregular obscuration of the star by circumstellar material. The observed increases of the star's brightness probably are caused by the existence of hot spots on the stellar surface.

We carried out a periodicity search using our data from 2010 May to 2015 September with \textsc{persea} Version 2.6 (written by G. Maciejewski on the $ANOVA$ technique, Schwarzenberg-Cherny (1996)) and with \textsc{period04} (Lenz \& Breger 2005) softwares. Our time-series analysis of the data indices a 2.450980 $\pm$ 0.029612 days period and led to the ephemeris:
\begin{equation}\label{eq1}
JD (Max) = 2455337.925020 + 2.450980 * E.
\end{equation}

False Alarm Probability estimation was done by randomly deleting about 10\%-15\% of the data for about 30 times and then retuning the period of determination. The period and starting age ($T_{0}$) determination remain stable, even with a subsample with 30\% of the data removed.
Fig. 15 exhibits the $V$-band folded light curve of LkH$\alpha$ 189 according to the ephemeris (1). Data obtained in $BRI$-bands show the same shape on periodicity. The found period is stable during time interval of several years and it is typical rotational period of young low-mass star. The periodicity could be caused by rotation modulation of the dark spots on the stellar surface.

\begin{figure}[]
  \begin{center}
    \centering{\epsfig{file=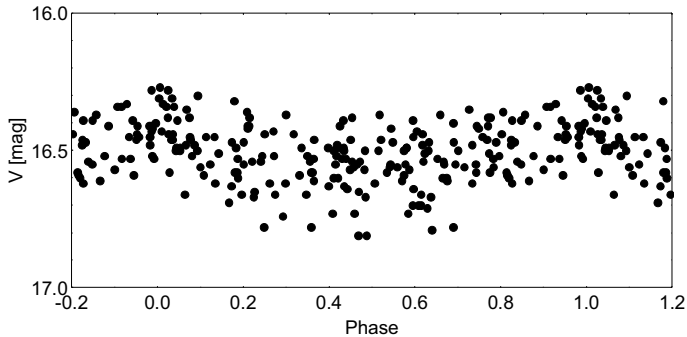, width=9.0cm}}
    \caption[]{$V$-band folded curve of LkH$\alpha$ 189}
    \label{fig15}
  \end{center}
\end{figure}


\subsection*{3.11. LkH$\alpha$ 191}

The star LkH$\alpha$ 191 was discovered and classified as a H$\alpha$ emission-line star by Herbig (1958) with photographic magnitude $m_{pg}$=14.00. Herbig \& Bell (1988) classified it as a T Tauri variable with photographic magnitude 12.08 and spectral type K0.
Terranegra et al. (1994) classified LkH$\alpha$ 191 as CTTS with $V$=12.95 mag and determined its photometric spectral type as K0V. Fernandez \& Eiroa (1996) presented the light curve in $V$-band of the star for the period 1988 July--1991 August. The light curve show approximately constant brightness (with variations in $V$-band from 12.89 to 12.99 mag).

Laugalys et al. (2006) measured $V$=13.06 mag, colour index $U-V$=3.506 mag and photometric spectral type K6e. Authors noticed the absence of a strong H$\alpha$ emission and presence of a small dust emission suggested no significant deviation of the star from the main sequence. The position of LkH$\alpha$ 191 in two-colour $J-H$/$H-K_{s}$ diagram is slightly below intrinsic T Tauri line. According to Laugalys et al. (2006) LkH$\alpha$ 191 has probably lost its envelope during the past 50 years.

Grankin et al. (2007) conducted long-term study of LkH$\alpha$ 191 during the period 1986 July--1995 August. The authors reported that the star's brightness in $V$ vary from 12.87 to 13.09 mag. 
Artemenko et al. (2012) determined the period of the star, which is found to be 2.08 days.


The $BVRI$ light curves of LkH$\alpha$ 191 from all our CCD observations (Poljan\v{c}i\'{c} Beljan et al. 2014; and the present paper) are shown in Fig. 16. The symbols used for the different telescopes are as in Fig. 3. The results of our recent multicolour CCD observations of the star are summarized in Table 14. The columns have the same contents as in Table 4.

\begin{figure}[]
  \begin{center}
    \centering{\epsfig{file=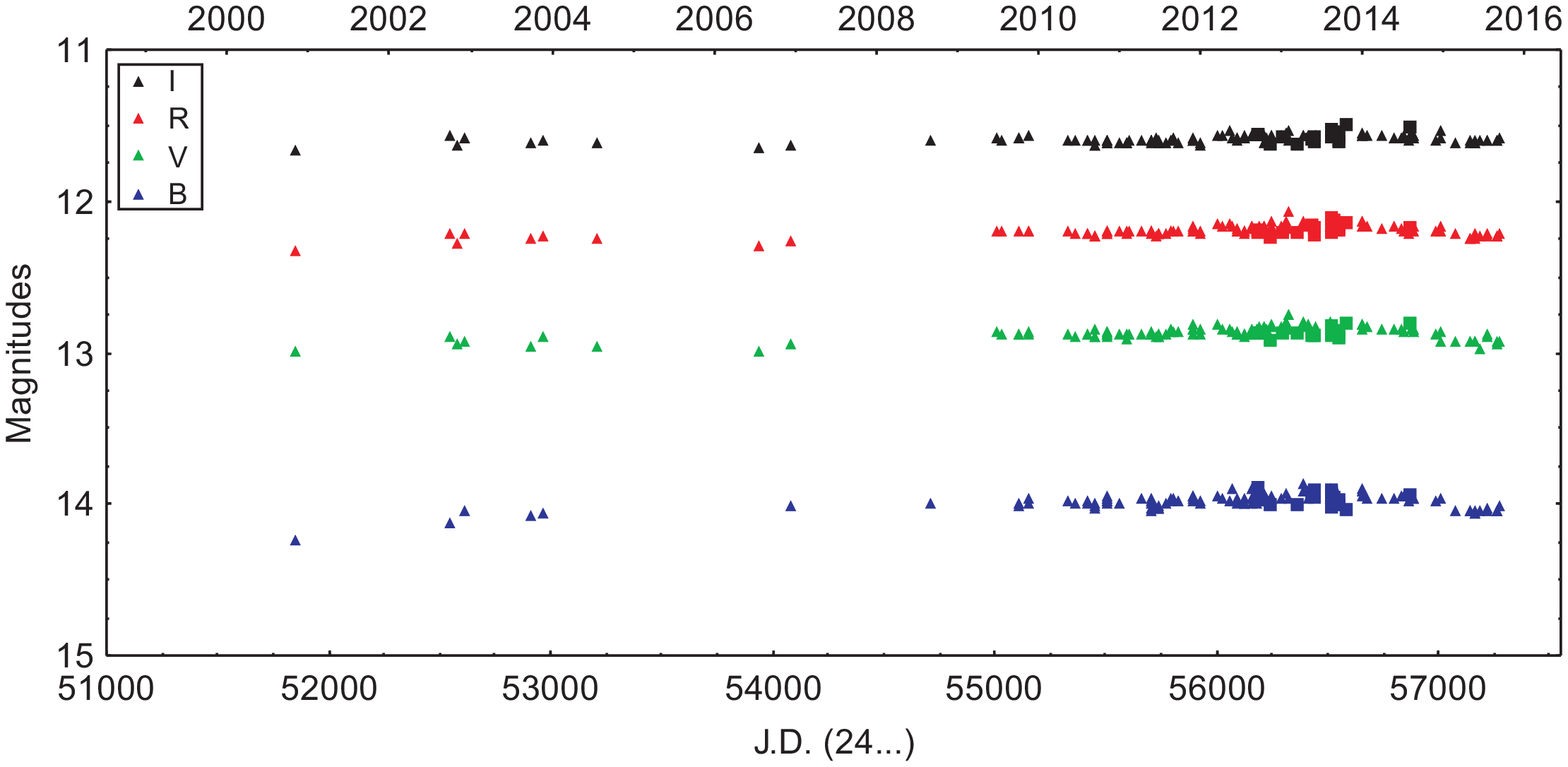, width=\textwidth}}
    \caption[]{CCD $IRVB$ light curves of LkH$\alpha$ 191 for the period 2000 October--2015 September}
    \label{fig16}
  \end{center}
\end{figure}

The brightness of LkH$\alpha$ 191 during the period 2000--2015 vary in the range 11.48--11.66 mag for $I$-band, 12.07--12.33 mag for $R$-band, 12.74--12.99 mag for $V$-band, and 13.87--14.23 mag for $B$-band. The observed amplitudes are $\Delta I$=0.18 mag, $\Delta R$=0.26 mag, $\Delta V$=0.25 mag, and $\Delta B$=0.36 mag.

It can be seen from Fig. 16 that LkH$\alpha$ 191 not shows some significant photometric activity. Our data agree with the hypothesis of an CTTS star, but LkH$\alpha$ 191 may also be post-T Tauri star.
We carried out a periodicity search using our data, but we could not confirm the findings of Artemenko et al. (2012).


\subsection*{3.12. [KW97] 53-11}

The star [KW97] 53-11 was mentioned in the catalog of Kohoutek \& Wehmeyer (1997, 1999) with brightness in the photovisual system $m_{pv}$=17.20 mag. Cohen \& Kuhi (1979) reported $V$=17.70 mag of the star and determined its spectral type as M2. Armond et al. (2011) report a lack of H$\alpha$ emission from this star.


The $RI$ light curves of [KW97] 53-11 from all our CCD observations (Poljan\v{c}i\'{c} Beljan et al. 2014; and the present paper) are shown in Fig. 17. The symbols used for the different telescopes are as in Fig. 3. The results of our recent multicolour CCD observations of the star are summarized in Table 15. The columns have the same contents as in Table 4.

\begin{figure}[]
  \begin{center}
    \centering{\epsfig{file=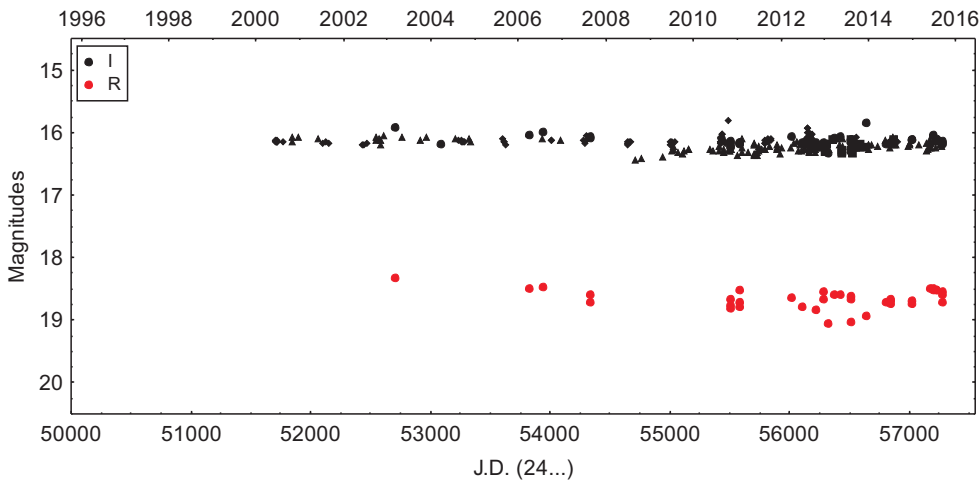, width=\textwidth}}
    \caption[]{CCD $IRVB$ light curves of [KW97] 53-11 for the period 1997 June--2015 September}
    \label{fig17}
  \end{center}
\end{figure}

The brightness of the star during the period 2000--2015 vary in the range 15.81--16.44 mag for $I$-band and 18.32--19.04 mag for $R$-band. The observed amplitudes are $\Delta I$=0.63 mag and $\Delta R$=0.72 mag in the same period. The magnitudes of the star in $V$ and $B$-band are below the limit of telescopes used.

It can be seen from Fig. 17 that during our study the brightness of [KW97] 53-11 vary around some intermediate level. The star shows both increases in the brightness with different amplitudes and short fading events in the brightness. Probably the observed fading events are caused by the existence of cool spots or group of spots, and the observed increases of the star's brightness likely are caused by the existence of hot spots on the stellar surface.
Evidences of periodicity in the brightness variability of [KW97] 53-11 are not detected.


\subsection*{3.13. [KW97] 53-23}

The star [KW97] 53-23 was mentioned in the catalog of Kohoutek \& Wehmeyer (1997, 1999) with only equatorial coordinates and finding charts available. Guieu et al. (2009) included the star in their list of YSO candidates and measured its magnitudes $I$=15.093, $V$=18.445, and $B$=20.649 mag.


The $BVRI$ light curves of [KW97] 53-23 from all our CCD observations (Poljan\v{c}i\'{c} Beljan et al. 2014; and the present paper) are shown in Fig. 18. The symbols used for the different telescopes are as in Fig. 3. The results of our recent multicolour CCD observations of the star are summarized in Table 16. The columns have the same contents as in Table 4.

\begin{figure}[]
  \begin{center}
    \centering{\epsfig{file=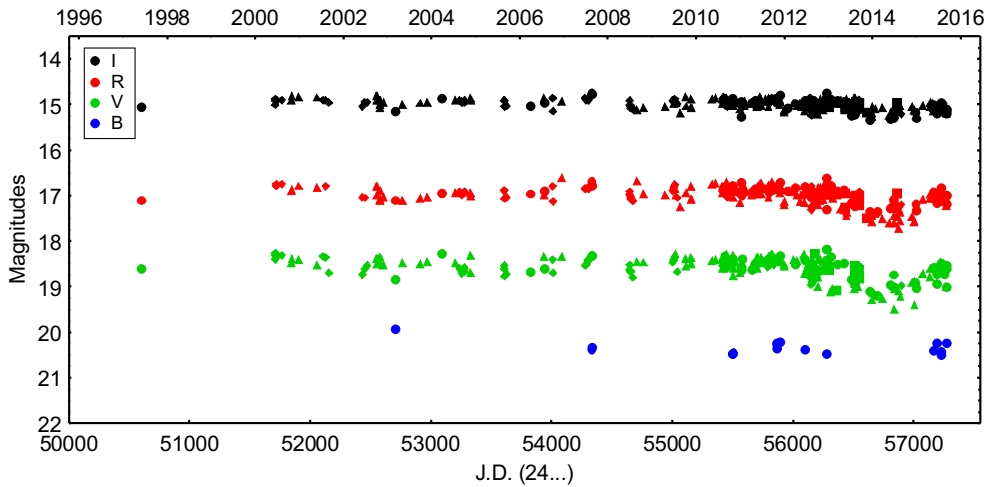, width=\textwidth}}
    \caption[]{CCD $IRVB$ light curves of [KW97] 53-23 for the period 1997 June--2015 September}
    \label{fig18}
  \end{center}
\end{figure}

It can be seen from Fig. 18 that [KW97] 53-23 shows both increases and drops in the brightness with different amplitudes. The brightness of the star during the period 1997--2015 vary in the range 14.74--15.34 mag for $I$-band, 16.28--17.73 mag for $R$-band, 18.01--19.49 mag for $V$-band, and 19.92--$~$20.49 mag for $B$-band. The observed amplitudes are $\Delta I$=0.60 mag, $\Delta R$=1.45 mag, $\Delta V$=1.48 mag, and $\Delta B$$~$0.57 mag. 

The deepest drop in the brightness of [KW97] 53-23 is observed during 2014. Due to the limit of our photometric data during the minimal brightness some magnitudes in $V$- and especially in $B$-band are not registered. We have only some $B$-band photometric data of the star collected with the 2-m RCC and the 1.3-m RC telescopes.

The observed increases in the brightness of [KW97] 53-23 are caused probably by the existence of hot spots on the stellar surface, typical of CTTS. The observed drops in the brightness are caused probably by the existence of cool spots on the stellar surface or by irregular obscuration of the star by circumstellar material.

On Fig. 19, the colour indices $V-I$ and $V-R$ versus $V$ magnitude during the period of our observations are plotted. The diagrams (especially $V/V-R$) shows evidences for bluing effect and probably the observed deepest drop in brightness are caused by obscuration of the star from circumstellar material. 

\begin{figure}
\begin{center}
\includegraphics[width=4cm]{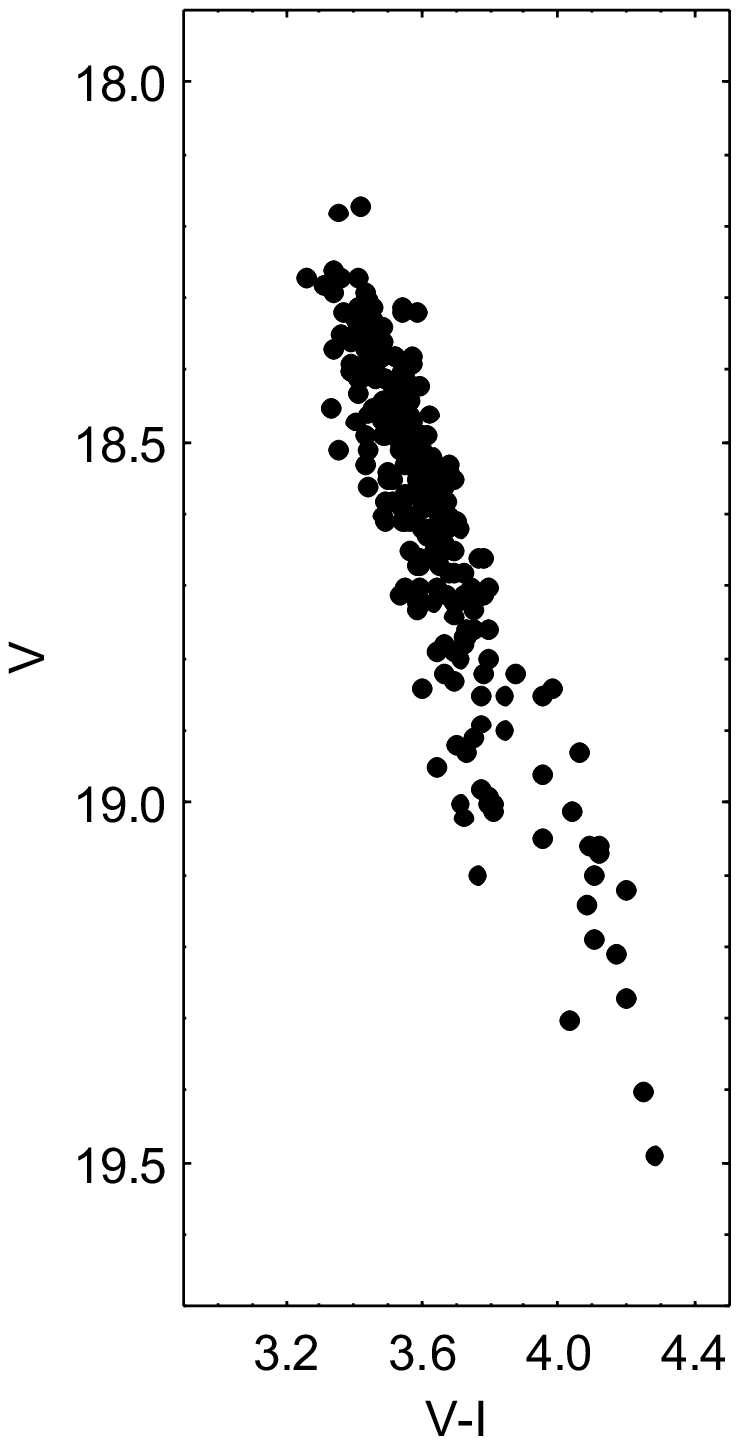}
\includegraphics[width=4cm]{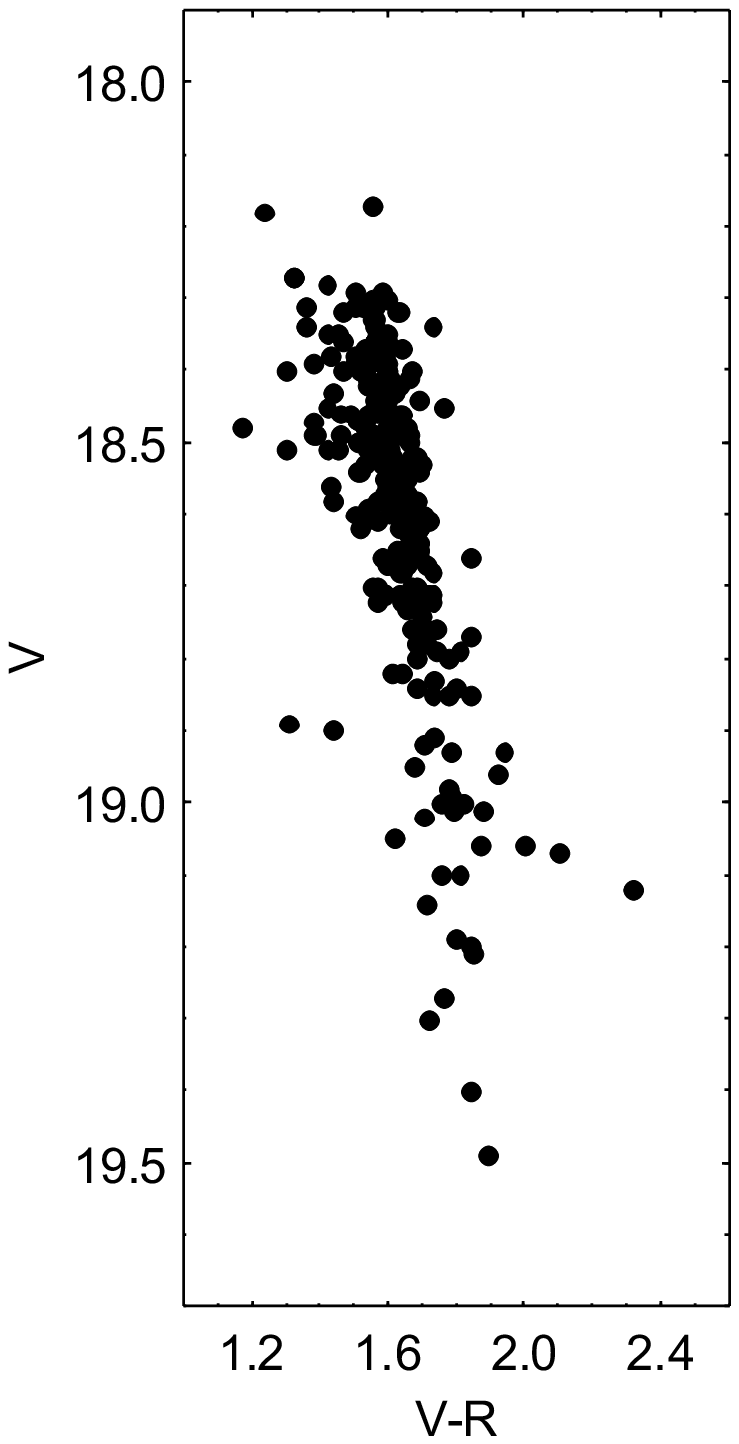}
\caption{Relationship between $V$ magnitude and the $V-I$ and $V-R$ colour indices of [KW97] 53-23 in the period 1997$-$2015}\label{fig19}
\end{center}
\end{figure}

In the two-colour diagram (Fig. 2) [KW97] 53-23 lies by 0.48 mag above the intrinsic T Tauri line. Therefore, the star have clear infrared excess, indicating the presence of a circumstellar disk. Evidences of periodicity in the brightness variability of the star are not detected.


\subsection*{3.14. [KW97] 53-36}

The star [KW97] 53-36 was mentioned in the catalog of Kohoutek \& Wehmeyer (1997, 1999) with it's brightness in the photographic system $m_{pg}$=12.70 mag.


The $BVRI$ light curves of [KW97] 53-36 from all our CCD observations (Poljan\v{c}i\'{c} Beljan et al. 2014; and the present paper) are shown in Fig. 20. The symbols used for the different telescopes are as in Fig. 3. The results of our recent multicolour CCD observations of the star are summarized in Table 17. The columns have the same contents as in Table 4.

\begin{figure}[]
  \begin{center}
    \centering{\epsfig{file=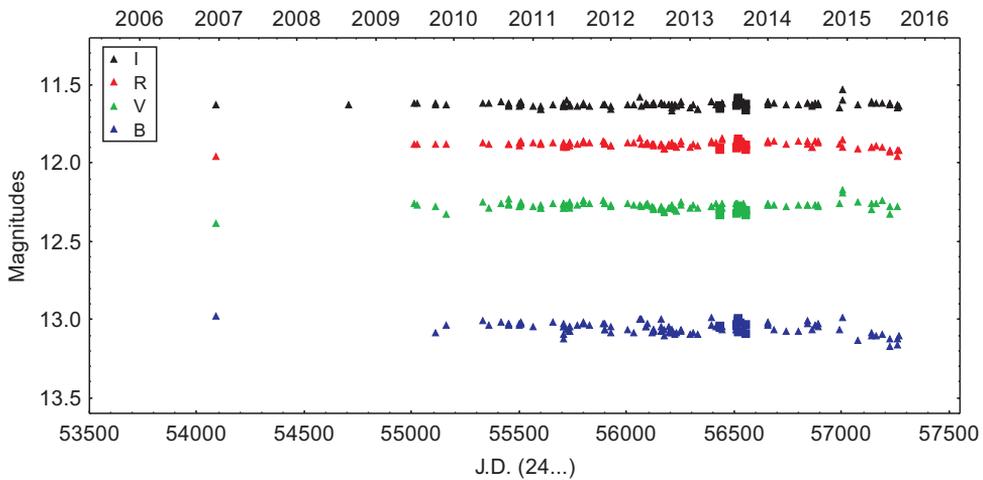, width=\textwidth}}
    \caption[]{CCD $IRVB$ light curves of [KW97] 53-36 for the period 2006 December--2015 September}
    \label{fig20}
  \end{center}
\end{figure}

It can be seen from Fig. 20 that [KW97] 53-36 not shows some significant photometric activity. The brightness of the star during the period 2006--2015 vary in the range 11.53--11.67 mag for $I$-band, 11.84--11.96 mag for $R$-band, 12.17--12.38 mag for $V$-band, and 12.98--13.17 mag for $B$-band. The observed amplitudes are $\Delta I$=0.14 mag, $\Delta R$=0.12 mag, $\Delta V$=0.21 mag, and $\Delta B$=0.19 mag in the same period.

In the two-colour diagram (Fig. 2) the star lies close to the intrinsic position of G0 stars with no infrared excess, i.e., there are no indications of the presence of a circumstellar disk. Most probably the star is a G giant with weak chromospheric activity. Evidences of periodicity in the brightness variability are not detected.

\section*{4. Conclusion}

We presented the light curves of 14 PMS stars located in NGC 7000/IC 5070 star-forming complex. Our results complement the previously rare insights into their photometry. Seven of studied stars (V752 Cyg, V1539 Cyg, V1716 Cyg, LkH$\alpha$ 186, LkH$\alpha$ 187, LkH$\alpha$ 191, [KW97] 53-23) shows characteristics for CTTS, one star (V1538 Cyg) probably is WTTS, for three objects (V1539 Cyg, LkH$\alpha$ 189, [KW97] 53-11) we waver between CTTS and WTTS, for their exact classification are necessary spectral observations, three stars (V1957 Cyg, V2051 Cyg, [KW97] 53-36) likely are evolved PMS stars, and one stars (V521 Cyg) from our study we found indications of UXor variability. For one investigated star, LkH$\alpha$ 189, we found periodicity -- 2.45-day period.

We are continuing to collect photometric observations of the field of 'Gulf of Mexico'. Further regular observations of the PMS stars in the same field will be of a great importance for their exact classification and are encouraging.


\section*{Acknowledgements}

This research has made use of the NASA's Astrophysics Data System Abstract Service, the SIMBAD database and the VizieR catalogue access tool, operated at CDS, Strasbourg, France. This publication makes use of data products from the Two Micron All Sky Survey, which is a joint project of the University of Massachusetts and the Infrared Processing and Analysis Center/California Institute of Technology, funded by the National Aeronautics and Space Administration and the National Science Foundation. The authors thank the Director of Skinakas Observatory Prof. I. Papamastorakis and Prof. I. Papadakis for the award of telescope time.

\newpage
\begin{appendix}

\section{ONLINE MATERIAL. Photometric CCD observations and data of the stars from our study}

{\small
\begin{longtable}{cccccccc}
\caption{Photometric CCD observations and data of V521 Cyg during the period 2014 August$-$2015 September}\\
\hline\hline
\noalign{\smallskip}  
Date & J.D. (24...) & I & R & V & B & Tel & CCD\\
\noalign{\smallskip}  
\hline
\endfirsthead
\caption{Continued.}\\
\hline\hline
\noalign{\smallskip}  
Date & J.D. (24...) & I & R & V & B & Tel & CCD\\
\noalign{\smallskip}  
\hline
\noalign{\smallskip}  
\endhead
\hline
\label{Tab4}
\endfoot
\noalign{\smallskip}
03.08.2014	&	56873.386	&	12.25	&	13.13	&	14.02	&	15.35	&	Sch	&	FLI	\\
04.08.2014	&	56874.407	&	12.27	&	13.11	&	13.97	&	15.33	&	Sch	&	FLI	\\
18.08.2014	&	56888.450	&	12.09	&	12.93	&	13.77	&	14.99	&	Sch	&	FLI	\\
19.08.2014	&	56889.365	&	12.04	&	12.85	&	13.68	&	14.93	&	Sch	&	FLI	\\
29.08.2014	&	56899.303	&	11.93	&	12.74	&	13.55	&	14.77	&	1.3-m	&	ANDOR	\\
26.11.2014	&	56988.181	&	11.95	&	12.74	&	13.53	&	14.74	&	Sch	&	FLI	\\
13.12.2014	&	57005.181	&	11.91	&	12.71	&	13.50	&	14.69	&	Sch	&	FLI	\\
14.12.2014	&	57006.220	&	11.94	&	12.73	&	13.51	&	-	&	Sch	&	FLI	\\
\hline \hline
\end{longtable}}

{\small
\begin{longtable}{cccccccc}
\caption{Photometric CCD observations and data of V752 Cyg during the period 2014 August$-$2015 September}\\
\hline\hline
\noalign{\smallskip}  
Date & J.D. (24...) & I & R & V & B & Tel & CCD\\
\noalign{\smallskip}  
\hline
\endfirsthead
\caption{Continued.}\\
\hline\hline
\noalign{\smallskip}  
Date & J.D. (24...) & I & R & V & B & Tel & CCD\\
\noalign{\smallskip}  
\hline
\noalign{\smallskip}  
\endhead
\hline
\label{Tab4}
\endfoot
\noalign{\smallskip}
03.08.2014	&	56873.386	&	15.10	&	16.37	&	17.43	&	18.36	&	Sch	&	FLI	\\
04.08.2014	&	56874.407	&	14.96	&	15.76	&	16.36	&	16.96	&	Sch	&	FLI	\\
18.08.2014	&	56888.450	&	15.11	&	15.98	&	16.60	&	17.30	&	Sch	&	FLI	\\
19.08.2014	&	56889.365	&	15.21	&	16.33	&	17.20	&	18.19	&	Sch	&	FLI	\\
26.11.2014	&	56988.181	&	15.07	&	16.06	&	16.70	&	17.62	&	Sch	&	FLI	\\
13.12.2014	&	57005.181	&	14.97	&	16.10	&	16.87	&	17.91	&	Sch	&	FLI	\\
14.12.2014	&	57006.220	&	15.21	&	16.43	&	17.16	&	-	&	Sch	&	FLI	\\
21.02.2015	&	57074.641	&	15.25	&	16.28	&	16.90	&	17.97	&	Sch	&	FLI	\\
23.04.2015	&	57136.487	&	15.21	&	16.60	&	17.47	&	18.77	&	Sch	&	FLI	\\
19.05.2015	&	57162.460	&	15.29	&	16.57	&	17.27	&	18.75	&	Sch	&	FLI	\\
12.06.2015	&	57186.444	&	14.96	&	15.89	&	16.50	&	17.39	&	Sch	&	FLI	\\
16.07.2015	&	57220.364	&	14.88	&	15.84	&	16.60	&	17.47	&	Sch	&	FLI	\\
17.07.2015	&	57221.417	&	14.97	&	15.96	&	16.68	&	17.50	&	Sch	&	FLI	\\
24.08.2015	&	57259.337	&	14.93	&	15.78	&	16.22	&	17.06	&	Sch	&	FLI	\\
25.08.2015	&	57260.342	&	15.30	&	16.30	&	16.91	&	17.95	&	Sch	&	FLI	\\
03.09.2015	& 57269.328	& 15.03	& 16.11	& 16.71	& 17.79	& Sch	& FLI \\
\hline \hline
\end{longtable}}

{\small
\begin{longtable}{cccccccc}
\caption{Photometric CCD observations and data of V1538 Cyg during the period 2013 May$-$2015 September}\\
\hline\hline
\noalign{\smallskip}  
Date & J.D. (24...) & I & R & V & B & Tel & CCD\\
\noalign{\smallskip}  
\hline
\endfirsthead
\caption{Continued.}\\
\hline\hline
\noalign{\smallskip}  
Date & J.D. (24...) & I & R & V & B & Tel & CCD\\
\noalign{\smallskip}  
\hline
\noalign{\smallskip}  
\endhead
\hline
\label{Tab4}
\endfoot
\noalign{\smallskip}
15.05.2013	&	56428.404	&	15.04	&	15.93	&	17.04	&	18.14	&	60-cm	&	FLI	\\
10.04.2013	&	56392.510	&	15.03	&	16.12	&	17.11	&	18.44	&	Sch	&	FLI	\\
11.04.2013	&	56394.482	&	15.00	&	15.99	&	16.94	&	18.40	&	Sch	&	FLI	\\
02.05.2013	&	56415.418	&	14.97	&	16.03	&	16.95	&	18.43	&	Sch	&	FLI	\\
17.05.2013	&	56430.408	&	15.00	&	15.95	&	16.96	&	-	&	60-cm	&	FLI	\\
19.05.2013	&	56432.405	&	14.98	&	15.99	&	16.81	&	-	&	60-cm	&	FLI	\\
30.05.2013	&	56443.367	&	14.96	&	16.00	&	16.97	&	18.50	&	Sch	&	FLI	\\
31.05.2013	&	56444.354	&	14.96	&	15.98	&	16.97	&	18.45	&	Sch	&	FLI	\\
\hline \hline
\end{longtable}}

{\small
\begin{longtable}{cccccccc}
\caption{Photometric CCD observations and data of V1539 Cyg during the period 2014 August$-$2015 September}\\
\hline\hline
\noalign{\smallskip}  
Date & J.D. (24...) & I & R & V & B & Tel & CCD\\
\noalign{\smallskip}  
\hline
\endfirsthead
\caption{Continued.}\\
\hline\hline
\noalign{\smallskip}  
Date & J.D. (24...) & I & R & V & B & Tel & CCD\\
\noalign{\smallskip}  
\hline
\noalign{\smallskip}  
\endhead
\hline
\label{Tab4}
\endfoot
\noalign{\smallskip}
03.08.2014	&	56873.386	&	13.57	&	14.59	&	15.65	&	17.01	&	Sch	&	FLI	\\
04.08.2014	&	56874.407	&	13.49	&	14.47	&	15.52	&	16.85	&	Sch	&	FLI	\\
18.08.2014	&	56888.450	&	13.61	&	14.66	&	15.74	&	17.13	&	Sch	&	FLI	\\
19.08.2014	&	56889.365	&	13.60	&	14.62	&	15.71	&	17.14	&	Sch	&	FLI	\\
29.08.2014	&	56899.303	&	13.58	&	14.61	&	15.66	&	17.04	&	1.3-m	&	ANDOR	\\
26.11.2014	&	56988.181	&	13.61	&	14.66	&	15.72	&	17.13	&	Sch	&	FLI	\\
13.12.2014	&	57005.181	&	13.54	&	14.57	&	15.54	&	16.98	&	Sch	&	FLI	\\
14.12.2014	&	57006.220	&	13.60	&	14.65	&	15.62	&	-	&	Sch	&	FLI	\\
\hline \hline
\end{longtable}}

{\small
\begin{longtable}{cccccccc}
\caption{Photometric CCD observations and data of V1716 Cyg during the period 2014 August$-$2015 September}\\
\hline\hline
\noalign{\smallskip}  
Date & J.D. (24...) & I & R & V & B & Tel & CCD\\
\noalign{\smallskip}  
\hline
\endfirsthead
\caption{Continued.}\\
\hline\hline
\noalign{\smallskip}  
Date & J.D. (24...) & I & R & V & B & Tel & CCD\\
\noalign{\smallskip}  
\hline
\noalign{\smallskip}  
\endhead
\hline
\label{Tab4}
\endfoot
\noalign{\smallskip}
03.08.2014	&	56873.386	&	15.09	&	16.48	&	17.67	&	19.29	&	Sch	&	FLI	\\
04.08.2014	&	56874.407	&	15.07	&	16.58	&	17.66	&	-	&	Sch	&	FLI	\\
18.08.2014	&	56888.450	&	15.14	&	16.59	&	17.82	&	-	&	Sch	&	FLI	\\
19.08.2014	&	56889.365	&	15.15	&	16.64	&	17.77	&	19.25	&	Sch	&	FLI	\\
29.08.2014	&	56899.303	&	14.99	&	16.41	&	17.59	&	19.21	&	1.3-m	&	ANDOR	\\
26.11.2014	&	56988.181	&	15.15	&	16.64	&	17.75	&	19.45	&	Sch	&	FLI	\\
13.12.2014	&	57005.181	&	15.09	&	16.64	&	17.64	&	-	&	Sch	&	FLI	\\
14.12.2014	&	57006.220	&	15.09	&	16.52	&	17.59	&	-	&	Sch	&	FLI	\\
\hline \hline
\end{longtable}}

{\small
\begin{longtable}{cccccccc}
\caption{Photometric CCD observations and data of V1957 Cyg during the period 2013 May$-$2015 September}\\
\hline\hline
\noalign{\smallskip}  
Date & J.D. (24...) & I & R & V & B & Tel & CCD\\
\noalign{\smallskip}  
\hline
\endfirsthead
\caption{Continued.}\\
\hline\hline
\noalign{\smallskip}  
Date & J.D. (24...) & I & R & V & B & Tel & CCD\\
\noalign{\smallskip}  
\hline
\noalign{\smallskip}  
\endhead
\hline
\label{Tab4}
\endfoot
\noalign{\smallskip}
15.05.2013	&	56428.404	&	14.65	&	15.41	&	16.31	&	17.52	&	60-cm	&	FLI	\\
10.04.2013	&	56392.510	&	14.57	&	15.45	&	16.32	&	-	&	Sch	&	FLI	\\
11.04.2013	&	56394.482	&	14.63	&	15.47	&	16.33	&	17.55	&	Sch	&	FLI	\\
02.05.2013	&	56415.418	&	14.63	&	15.47	&	16.33	&	17.60	&	Sch	&	FLI	\\
17.05.2013	&	56430.408	&	14.65	&	15.52	&	16.29	&	-	&	60-cm	&	FLI	\\
19.05.2013	&	56432.405	&	14.62	&	15.40	&	16.26	&	17.39	&	60-cm	&	FLI	\\
30.05.2013	&	56443.367	&	14.57	&	15.44	&	16.30	&	17.50	&	Sch	&	FLI	\\
31.05.2013	&	56444.354	&	11.57	&	15.43	&	16.26	&	17.47	&	Sch	&	FLI	\\
\hline \hline
\end{longtable}}

{\small
\begin{longtable}{cccccccc}
\caption{Photometric CCD observations and data of V2051 Cyg during the period 2014 August$-$2015 September}\\
\hline\hline
\noalign{\smallskip}  
Date & J.D. (24...) & I & R & V & B & Tel & CCD\\
\noalign{\smallskip}  
\hline
\endfirsthead
\caption{Continued.}\\
\hline\hline
\noalign{\smallskip}  
Date & J.D. (24...) & I & R & V & B & Tel & CCD\\
\noalign{\smallskip}  
\hline
\noalign{\smallskip}  
\endhead
\hline
\label{Tab4}
\endfoot
\noalign{\smallskip}
03.08.2014	&	56873.386	&	13.93	&	15.65	&	16.58	&	18.12	&	Sch	&	FLI	\\
04.08.2014	&	56874.407	&	13.92	&	15.59	&	16.69	&	18.08	&	Sch	&	FLI	\\
18.08.2014	&	56888.450	&	13.94	&	15.58	&	16.60	&	18.10	&	Sch	&	FLI	\\
19.08.2014	&	56889.365	&	13.93	&	15.57	&	16.62	&	18.15	&	Sch	&	FLI	\\
29.08.2014	&	56899.303	&	13.92	&	15.52	&	16.58	&	18.11	&	1.3-m	&	ANDOR	\\
26.11.2014	&	56988.181	&	13.93	&	15.57	&	16.59	&	18.12	&	Sch	&	FLI	\\
13.12.2014	&	57005.181	&	13.94	&	15.55	&	16.59	&	18.09	&	Sch	&	FLI	\\
14.12.2014	&	57006.220	&	13.93	&	15.53	&	16.48	&	-	&	Sch	&	FLI	\\
\hline \hline
\end{longtable}}

{\small
\begin{longtable}{cccccccc}
\caption{Photometric CCD observations and data of LkH$\alpha$ 186 during the period 2013 April$-$2015 September}\\
\hline\hline
\noalign{\smallskip}  
Date & J.D. (24...) & I & R & V & B & Tel & CCD\\
\noalign{\smallskip}  
\hline
\endfirsthead
\caption{Continued.}\\
\hline\hline
\noalign{\smallskip}  
Date & J.D. (24...) & I & R & V & B & Tel & CCD\\
\noalign{\smallskip}  
\hline
\noalign{\smallskip}  
\endhead
\hline
\label{Tab4}
\endfoot
\noalign{\smallskip}
10.04.2013	&	56392.510	&	14.44	&	15.99	&	17.10	&	-	&	Sch	&	FLI	\\
11.04.2013	&	56394.482	&	14.46	&	16.02	&	17.19	&	18.68	&	Sch	&	FLI	\\
02.05.2013	&	56415.418	&	14.59	&	16.19	&	17.40	&	19.12	&	Sch	&	FLI	\\
15.05.2013	&	56428.404	&	14.54	&	15.98	&	17.22	&	-	&	60-cm	&	FLI	\\
17.05.2013	&	56430.408	&	14.54	&	16.07	&	17.26	&	-	&	60-cm	&	FLI	\\
19.05.2013	&	56432.405	&	14.65	&	16.16	&	17.48	&	-	&	60-cm	&	FLI	\\
30.05.2013	&	56443.367	&	14.58	&	16.19	&	17.38	&	19.20	&	Sch	&	FLI	\\
31.05.2013	&	56444.354	&	14.55	&	16.18	&	17.45	&	-	&	Sch	&	FLI	\\
\hline \hline
\end{longtable}}

{\small
\begin{longtable}{cccccccc}
\caption{Photometric CCD observations and data of LkH$\alpha$ 187 during the period 2013 April$-$2015 September}\\
\hline\hline
\noalign{\smallskip}  
Date & J.D. (24...) & I & R & V & B & Tel & CCD\\
\noalign{\smallskip}  
\hline
\endfirsthead
\caption{Continued.}\\
\hline\hline
\noalign{\smallskip}  
Date & J.D. (24...) & I & R & V & B & Tel & CCD\\
\noalign{\smallskip}  
\hline
\noalign{\smallskip}  
\endhead
\hline
\label{Tab4}
\endfoot
\noalign{\smallskip}
10.04.2013	&	56392.510	&	14.80	&	16.43	&	17.81	&	-	&	Sch	&	FLI	\\
11.04.2013	&	56394.482	&	14.76	&	16.33	&	17.71	&	19.23	&	Sch	&	FLI	\\
02.05.2013	&	56415.418	&	14.73	&	16.30	&	17.71	&	19.33	&	Sch	&	FLI	\\
15.05.2013	&	56428.404	&	14.63	&	16.14	&	17.43	&	-	&	60-cm	&	FLI	\\
17.05.2013	&	56430.408	&	14.63	&	16.15	&	17.31	&	-	&	60-cm	&	FLI	\\
19.05.2013	&	56432.405	&	14.74	&	16.30	&	17.40	&	-	&	60-cm	&	FLI	\\
30.05.2013	&	56443.367	&	14.70	&	16.18	&	17.70	&	-	&	Sch	&	FLI	\\
31.05.2013	&	56444.354	&	14.71	&	16.28	&	17.78	&	-	&	Sch	&	FLI	\\
\hline \hline
\end{longtable}}

{\small
\begin{longtable}{cccccccc}
\caption{Photometric CCD observations and data of LkH$\alpha$ 189 during the period 2013 May$-$2015 September}\\
\hline\hline
\noalign{\smallskip}  
Date & J.D. (24...) & I & R & V & B & Tel & CCD\\
\noalign{\smallskip}  
\hline
\endfirsthead
\caption{Continued.}\\
\hline\hline
\noalign{\smallskip}  
Date & J.D. (24...) & I & R & V & B & Tel & CCD\\
\noalign{\smallskip}  
\hline
\noalign{\smallskip}  
\endhead
\hline
\label{Tab4}
\endfoot
\noalign{\smallskip}
15.05.2013	&	56428.404	&	13.82	&	15.12	&	16.34	&	17.87	&	60-cm	&	FLI	\\
10.04.2013	&	56392.510	&	13.93	&	15.23	&	16.52	&	-	&	Sch	&	FLI	\\
11.04.2013	&	56394.482	&	13.89	&	15.24	&	16.45	&	18.01	&	Sch	&	FLI	\\
02.05.2013	&	56415.418	&	13.94	&	15.25	&	16.48	&	17.98	&	Sch	&	FLI	\\
17.05.2013	&	56430.408	&	13.90	&	15.23	&	16.48	&	-	&	60-cm	&	FLI	\\
19.05.2013	&	56432.405	&	13.92	&	15.22	&	16.47	&	17.96	&	60-cm	&	FLI	\\
30.05.2013	&	56443.367	&	13.85	&	15.16	&	16.34	&	18.16	&	Sch	&	FLI	\\
31.05.2013	&	56444.354	&	13.82	&	15.10	&	16.28	&	17.90	&	Sch	&	FLI	\\
\hline \hline
\end{longtable}}

{\small
\begin{longtable}{cccccccc}
\caption{Photometric CCD observations and data of LkH$\alpha$ 191 during the period 2013 April$-$2015 September}\\
\hline\hline
\noalign{\smallskip}  
Date & J.D. (24...) & I & R & V & B & Tel & CCD\\
\noalign{\smallskip}  
\hline
\endfirsthead
\caption{Continued.}\\
\hline\hline
\noalign{\smallskip}  
Date & J.D. (24...) & I & R & V & B & Tel & CCD\\
\noalign{\smallskip}  
\hline
\noalign{\smallskip}  
\endhead
\hline
\label{Tab4}
\endfoot
\noalign{\smallskip}
10.04.2013	&	56392.510	&	11.56	&	12.14	&	12.79	&	13.87	&	Sch	&	FLI	\\
11.04.2013	&	56394.482	&	11.56	&	12.16	&	12.83	&	13.91	&	Sch	&	FLI	\\
02.05.2013	&	56415.418	&	11.56	&	12.17	&	12.82	&	13.96	&	Sch	&	FLI	\\
15.05.2013	&	56428.404	&	11.58	&	12.15	&	12.87	&	13.92	&	60-cm	&	FLI	\\
17.05.2013	&	56430.408	&	11.59	&	12.21	&	12.88	&	13.94	&	60-cm	&	FLI	\\
19.05.2013	&	56432.405	&	11.56	&	12.17	&	12.87	&	13.90	&	60-cm	&	FLI	\\
30.05.2013	&	56443.367	&	11.57	&	12.16	&	12.84	&	13.94	&	Sch	&	FLI	\\
31.05.2013	&	56444.354	&	11.57	&	12.15	&	12.83	&	13.94	&	Sch	&	FLI	\\
\hline \hline
\end{longtable}}

{\small
\begin{longtable}{cccccc}
\caption{Photometric CCD observations and data of [KW97] 53-11 during the period 2013 April$-$2015 September}\\
\hline\hline
\noalign{\smallskip}  
Date & J.D. (24...) & I & R & Tel & CCD\\
\noalign{\smallskip}  
\hline
\endfirsthead
\caption{Continued.}\\
\hline\hline
\noalign{\smallskip}  
Date & J.D. (24...) & I & R & Tel & CCD\\
\noalign{\smallskip}  
\hline
\noalign{\smallskip}  
\endhead
\hline
\label{Tab4}
\endfoot
\noalign{\smallskip}
10.04.2013	&	56392.510	&	16.11	&	-	&	Sch	&	FLI	\\
11.04.2013	&	56394.482	&	16.09	&	-	&	Sch	&	FLI	\\
02.05.2013	&	56415.418	&	16.19	&	-	&	Sch	&	FLI	\\
15.05.2013	&	56428.404	&	16.26	&	-	&	60-cm	&	FLI	\\
17.05.2013	&	56430.408	&	16.31	&	-	&	60-cm	&	FLI	\\
19.05.2013	&	56432.405	&	16.30	&	-	&	60-cm	&	FLI	\\
30.05.2013	&	56443.367	&	16.10	&	-	&	Sch	&	FLI	\\
31.05.2013	&	56444.354	&	16.22	&	-	&	Sch	&	FLI	\\
\hline \hline
\end{longtable}}

{\small
\begin{longtable}{cccccccc}
\caption{Photometric CCD observations and data of [KW97] 53-23 during the period 2013 April$-$2015 September}\\
\hline\hline
\noalign{\smallskip}  
Date & J.D. (24...) & I & R & V & B & Tel & CCD\\
\noalign{\smallskip}  
\hline
\endfirsthead
\caption{Continued.}\\
\hline\hline
\noalign{\smallskip}  
Date & J.D. (24...) & I & R & V & B & Tel & CCD\\
\noalign{\smallskip}  
\hline
\noalign{\smallskip}  
\endhead
\hline
\label{Tab4}
\endfoot
\noalign{\smallskip}
10.04.2013	&	56392.510	&	14.98	&	-	&	-	&	-	&	Sch	&	FLI	\\
11.04.2013	&	56394.482	&	14.92	&	17.30	&	-	&	-	&	Sch	&	FLI	\\
02.05.2013	&	56415.418	&	14.99	&	17.29	&	-	&	-	&	Sch	&	FLI	\\
15.05.2013	&	56428.404	&	15.05	&	17.25	&	-	&	-	&	60-cm	&	FLI	\\
17.05.2013	&	56430.408	&	14.95	&	17.30	&	-	&	-	&	60-cm	&	FLI	\\
19.05.2013	&	56432.405	&	15.10	&	16.95	&	-	&	-	&	60-cm	&	FLI	\\
30.05.2013	&	56443.367	&	14.98	&	17.43	&	-	&	-	&	Sch	&	FLI	\\
31.05.2013	&	56444.354	&	14.88	&	17.36	&	-	&	-	&	Sch	&	FLI	\\
\hline \hline
\end{longtable}}

{\small
\begin{longtable}{cccccccc}
\caption{Photometric CCD observations and data of [KW97] 53-36 during the period 2013 April$-$2015 September}\\
\hline\hline
\noalign{\smallskip}  
Date & J.D. (24...) & I & R & V & B & Tel & CCD\\
\noalign{\smallskip}  
\hline
\endfirsthead
\caption{Continued.}\\
\hline\hline
\noalign{\smallskip}  
Date & J.D. (24...) & I & R & V & B & Tel & CCD\\
\noalign{\smallskip}  
\hline
\noalign{\smallskip}  
\endhead
\hline
\label{Tab4}
\endfoot
\noalign{\smallskip}
10.04.2013	&	56392.510	&	11.61	&	11.86	&	12.28	&	12.99	&	Sch	&	FLI	\\
11.04.2013	&	56394.482	&	11.61	&	11.87	&	12.28	&	13.04	&	Sch	&	FLI	\\
02.05.2013	&	56415.418	&	11.63	&	11.87	&	12.26	&	13.05	&	Sch	&	FLI	\\
15.05.2013	&	56428.404	&	11.63	&	11.90	&	12.29	&	13.04	&	60-cm	&	FLI	\\
17.05.2013	&	56430.408	&	11.65	&	11.91	&	12.33	&	13.05	&	60-cm	&	FLI	\\
30.05.2013	&	56443.367	&	11.62	&	11.84	&	12.29	&	13.06	&	Sch	&	FLI	\\
31.05.2013	&	56444.354	&	11.62	&	11.85	&	12.26	&	13.05	&	Sch	&	FLI	\\
04.08.2013	&	56509.288	&	11.61	&	11.86	&	12.29	&	13.05	&	Sch	&	FLI	\\
\hline \hline
\end{longtable}}

\end{appendix}


\begin{thebibliography}{}

\bibitem{}
Armond, T., Reipurth, B., Bally, J., Aspin, C., 2011, {\em A\&A, 528, A125}

\bibitem{}
Artemenko, S. A., Grankin, K. N., Petrov, P., 2012, {\em Astronomy Letters, 38, 783}

\bibitem{}
Bally, J., Ginsburg, A., Probst, R., Reipurth, B., Shirley, Y. L., Stringfellow, G. S., 2014, {\em AJ, 148, 120}

\bibitem{}
Bertout, C., 1989, {\em ARA\&A, 27, 3515}

\bibitem{}
Bessell, M. S., Brett, J. M., 1988, {\em PASP, 100, 1134}

\bibitem{}
Bibo, E. A., Th\'{e}, P. S., 1990, {\em A\&A, 236, 155}

\bibitem{}
Carpenter, J. M., 2001, {\em AJ, 121, 2851}

\bibitem{}
Chavarr\'{i}a-K., C., Terranegra, L., Alcal\'{a}, J. M., Neri, L., 1989, {\em RMxAA, 18, 178}

\bibitem{}
Chavushian, H. S., Jankovics, I., 1985, {\em IBVS, 2814, 1}

\bibitem{}
Cohen, M., Kuhi, L. V., 1979, {\em ApJ Suppl., 41, 743}

\bibitem{}
Corbally, C. J., Strai\v{z}ys, V., Laugalys, V., 2009, {\em Balt. Astron., 18, 111}

\bibitem{}
Dullemond, C. P., van den Ancker, M. E., Acke, B., van Boekel, R., 2003, {\em ApJ, 594, L47}

\bibitem{}
Erastova, L. K., Tsvetkov, M. K., 1974, {\em IBVS, 909, 1}

\bibitem{}
Erastova, L. K., Tsvetkov, M. K., 1978, {\em Perem. Zvezdy, Byull., 21, 79}

\bibitem{}
Fernandez, M., Eiroa, C., 1996, {\em A\&A, 310, 143}

\bibitem{}
Filin, A. Ya., 1974, {\em Perem. Zvezdy, Prolozh., 2, 63}

\bibitem{}
Findeisen, K., Hillenbrand, L., Ofek, E., Levitan, D., Sesar, B., Laher, R., Surace, J., 2013, {\em ApJ, 768, 93}

\bibitem{}
Gieseking, F., Schumann, J. D., 1976, {\em A\&AS, 26, 367}

\bibitem{}
Grankin, K. N., Melnikov, S. Yu., Bouvier, J., Herbst, W., Shevchenko, V. S., 2007, {\em A\&AS, 461, 183}

\bibitem{}
Grinin, V. P., Kiselev, N. N., Minikulov, N. Kh., Chernova, G. P., Voshchinnikov, N. V., 1991, {\em Ap\&SS, 186, 283}

\bibitem{}
Guieu, S., Rebull, L. M., Stauffer, J. R., Hillenbrand, L. A., Carpenter, J. M. et al., 2009, {\em ApJ, 697, 787}

\bibitem{}
Herbig, G. H., 1958, {\em ApJ, 128, 259}

\bibitem{}
Herbig, G. H., Bell, K. R., 1988, {\em Third catalog of emission-line stars of the Orion population, Lick Observatory Bulletin 1111, Santa Cruz: Lick Observatory, 90}

\bibitem{}
Herbst, W., Shevchenko, V. S., 1999, {\em AJ, 118, 1043}

\bibitem{}
Herbst, W., Eisloffel, J., Mundt, R., Scholz, A., 2007, {\em Protostars and Planets V, Ed. B. Reipurth, D. Jewitt and K. Keil, 297}

\bibitem{}
Hoffmeister, C., 1949, {\em AN, 278, 24}

\bibitem{}
Ibryamov, S. I., Semkov, E. H., Peneva, S. P., 2015a, {\em Bulgarian Astronomical Journal, 22, 3}

\bibitem{}
Ibryamov, S. I., Semkov, E. H., Peneva, S. P., 2015b, {\em PASA, 32, e021}

\bibitem{}
Joy, A. H., 1945, {\em ApJ, 102, 168}

\bibitem{}
Kohoutek, L., Wehmeyer, R., 1997, {\em Abhandlungen aus der Hamburger Sternwarte, Band 11, Teil 1, 2}

\bibitem{}
Kohoutek, L., Wehmeyer, R., 1999, {\em A\&AS, 134, 255}

\bibitem{}
Laugalys, V., Strai\v{z}ys, V., 2002, {\em Balt. Astr., 11, 205}

\bibitem{}
Laugalys, V., Strai\v{z}ys, V., Vrba, F. J., Boyle, R. P., Philip, A. G. D., Kazlauskas, A., 2006, {\em Balt. Astron., 15, 483}

\bibitem{}
Lenz, P., Breger, M., 2005, {\em CoAst, 146, 53}

\bibitem{}
M\'{e}nard, F., Bertout, C., 1999, {\em The Origin of Stars and Planetary Systems, Ed. V. J. Lada and N. D. Kylafis, 341}

\bibitem{}
Meyer, M. R., Calvet, N., Hillenbrand, L. A., 1997, {\em AJ, 114, 288}

\bibitem{}
Miller, A. A., Hillenbrand, L. A., Covey, K. R., Poznanski, D., Silverman, J. M. et al., 2011, {\em ApJ, 730, 80}

\bibitem{}
Mitskevich, M. P., Pavlenko, E. P., 2001, {\em Ap, 44, 411}

\bibitem{}
Parsamian, E. S., Chavira, E., Gonzales, G., 1994, {\em IBVS, 4046, 1}

\bibitem{}
Poljan\v{c}i\'{c} Beljan, I., Jurdana-\v{S}epi\'{c}, R., Semkov, E., Ibryamov, S., Peneva, S., Tsvetkov, M., 2014, {\em A\&A, 568, A49}

\bibitem{}
Samus, N. N., Durlevich, O. V. et al., 2009, {\em VizieR Online Data Catalog: General Catalogue of Variable Stars, 1, 2025}

\bibitem{}
Schwarzenberg-Cherny, A., 1996, {\em ApJ, 460, 107}

\bibitem{}
Semkov, E. H., Peneva, S. P., Munari, U., Milani, A., Valisa, P., 2010, {\em A\&A, 523, L3}

\bibitem{}
Semkov, E. H., Peneva, S. P., Munari, U., Tsvetkov, M. K., Jurdana-S\v{e}pi\'{c}, R., de Miguel, E., Schwartz, R. D., Dimitrov, D. P., Kjurkchieva, D. P., Radeva, V. S., 2012, {\em A\&A, 542, A43}

\bibitem{}
Semkov, E. H., Peneva, S. P., Ibryamov, S. I., Dimitrov, D. P., 2014, {\em Bulgarian Astronomical Journal, 20, 59}

\bibitem{}
Skrutskie, M. F., Cutri, R. M., Stiening, R., Weinberg, M. D., Schneider, S. et al., 2006, {\em AJ, 131, 1163}

\bibitem{}
Strai\v{z}ys, V., Corbally, C. J., Laugalys, V., 2008, {\em Balt. Astron., 17, 125}

\bibitem{}
Terranegra, L., Chavarr\'{i}a-K., C., Diaz, S., Gonzales-Patino, D., 1994, {\em A\&AS, 104, 557}

\bibitem{}
Voshchinnikov, N. V., 1989, {\em Astrofizika, 30, 509}

\bibitem{}
Weintraub, D. A., 1990, {\em ApJS, 74, 575}

\bibitem{}
Welin, G., 1973, {\em A\&AS, 9, 183}

\end{thebibliography}
\end{document}